\documentclass[acmsmall,nonacm]{acmart}
\usepackage[english]{babel}
\usepackage{blindtext}
\usepackage{caption, subcaption}
\usepackage{arydshln}
\usepackage{algorithm}
\usepackage{algpseudocode}
\usepackage[normalem]{ulem}
\usepackage{enumitem}

\algrenewcommand\algorithmicrequire{\textbf{Input:}}
\algrenewcommand\algorithmicensure{\textbf{Output:}}

\begin{document}
\title{Rate-Fidelity Control for Wide-Area Quantum Links}

\author{Connor Clayton}
\email{cbclayto@cs.umd.edu}
\affiliation{%
  \institution{University of Maryland}
  \city{College Park} 
  \state{Maryland} 
  \country{USA}
}

\author{Cory Nunn}
\email{cory.nunn@nist.gov}
\affiliation{%
  \institution{National Institute of Standards and Technology}
  \city{Gaithersburg} 
  \state{Maryland} 
  \country{USA}
}
\affiliation{%
  \institution{University of Maryland}
  \city{College Park} 
  \state{Maryland} 
  \country{USA}
}

\author{Quinn Carmack}
\email{qcarmack@terpmail.umd.edu}
\affiliation{%
  \institution{University of Maryland}
  \city{College Park} 
  \state{Maryland} 
  \country{USA}
}

\author{Wayne McKenzie}
\email{wmckenzie@ltsnet.net}
\affiliation{%
  \institution{Laboratory for Telecommunication Sciences}
  \city{College Park} 
  \state{Maryland} 
  \country{USA}
}

\author{Anne Marie Richards}
\email{arichards@ltsnet.net}
\affiliation{%
  \institution{Laboratory for Telecommunication Sciences}
  \city{College Park} 
  \state{Maryland} 
  \country{USA}
}

\author{Xiaodi Wu}
\email{xwu@cs.umd.edu}
\affiliation{%
  \institution{University of Maryland}
  \city{College Park} 
  \state{Maryland} 
  \country{USA}
}

\author{Bobby Bhattacharjee}
\email{bobby@cs.umd.edu}
\affiliation{%
  \institution{University of Maryland}
  \city{College Park} 
  \state{Maryland} 
  \country{USA}
}

\renewcommand{\shortauthors}{Connor Clayton et al.}
\newcommand{\Fmin}{F_\text{min}\xspace}
\newcommand{\Fsource}{F_\text{sd}\xspace}
\newcommand{\Ftrans}{F_\text{pol}\xspace}
\newcommand{\Ftrigger}{F_\text{trigger}\xspace}
\newcommand{\Ftarget}{F_\text{target}\xspace}
\newcommand{\Teff}{T_\text{eff}\xspace}
\newcommand{\timeout}{T_\text{timeout}\xspace}

\newcommand{\mytilde}{\raisebox{0.5ex}{\texttildelow}}

\begin{abstract}

Quantum network links must distribute entanglement at high rates while satisfying application-specified fidelity demands. However, wide-area deployed fiber links suffer from polarization drift which destabilizes end-to-end fidelity and forces periodic compensation. Current deployments often use active stabilization with fixed control policies, and improvements generally stem from advances in quantum hardware. Meanwhile, software control remains relatively underexplored.

Here, we formulate quantum link operation as a joint control problem over tunable rate-fidelity tradeoffs and uncontrollable link drift. From this framework, we construct a link control protocol that dynamically adapts source pump power and polarization compensation to maximize entanglement distribution rate subject to a minimum fidelity constraint. We evaluate the protocol through trace-driven simulations driven by data from a 64 km deployed optical fiber. Compared with optimized static policies, our adaptive controller improves mean entanglement distribution rate by 14\% over a 24 hour trace, without requiring any offline policy optimization. Our results show that software-based physical layer control can provide a practical mechanism for improving near-term quantum link performance without requiring additional quantum hardware.

\end{abstract}

\maketitle

\section{Introduction}

Quantum networks promise to provide revolutionary capabilities in
areas such as secure communications~\cite{bb84, ekert1991quantum},
distributed quantum sensing~\cite{quantum-sensing1,
  gottesman-long-baseline-telemetry}, and modular quantum
computing~\cite{distributed-qc, modular-qc}.  To realize the utility
of these technologies, quantum network architectures and protocols
must scale both in terms of geographic distance and quality of
information transfer~\cite{wehner-road-ahead}.

The scaling of these metrics are at odds with each other.  Information
in a quantum network exists in the form of quantum bits (qubits) whose
signals are necessarily weak, with each transmitted qubit generally
consisting of a single photon.  Quantum communication is therefore
highly susceptible to loss and noise, both of which scale unfavorably
with increasing distance.

Most applications of quantum networks require the distribution of high-quality \emph{entanglement} between remote parties.
Entanglement is a property of two or more qubits that allows for correlations even across large distances.
The quality of an entangled quantum system can be measured by its \emph{fidelity} $F$, where $0\leq F \leq 1$ and $F=1$ indicates that the system is in the ``ideal'' target state. 
In practice, system imperfections and environmental noise cause $F$ to be less than $1$; however, such noisy states are often sufficient and a higher-fidelity state may be attained by consuming many noisy copies through a process known as entanglement distillation~\cite{horodecki-entanglement-distillation}.

Significant efforts are underway to design, develop, test, and deploy
hardware components which will enable high-fidelity transmission of entanglement~\cite{li_review_2023, boston-qnet, lukin-telecom-memory-entanglement}.  Concurrently,
considerable effort has been put into developments at higher layers of
the quantum protocol stack~\cite{quantum-protocol-stack} which manage
resources to deliver entanglement in complex networks~\cite{Q-CAST, multi-partite-routing, quarc, link-layer-protocol, qnet-os}.
However, there is a gap in the current research landscape: most
protocols for wide-area quantum networks generally assume high-volume,
consistent-fidelity entanglement generation --- conditions that the
physical layer is not expected to meet in the near- to
medium-term~\cite{li_review_2023}.

We propose a framework for considering optimization problems relevant
to the performance of \emph{present-day quantum networks}.  The family
of optimization problems we define here accounts for both
\emph{controllable parameters} of quantum link hardware, which admit
tradeoffs between fidelity and performance metrics like entanglement distribution rate, and
\emph{uncontrollable parameters} that diminish link fidelity, such as
polarization drift.
Neither set of parameters has been systematically analyzed
  for wide-area links, especially in conjunction. 
We introduce an
optimization problem that considers both sets of parameters
simultaneously, and present a link control protocol that tunes the
controllable parameters and adapts to the uncontrollable to achieve
high entanglement distribution rates over a wide-area link.

Our adaptive control protocol takes as input an end-to-end entanglement fidelity target,
and aims
to maximize the entanglement distribution rate over the link.  Absent link degradation, the entanglement rate can be easily maximized
with a proper selection of controllable parameters.
However, in a realistic wide-area setting, physical properties 
and environmental factors
(e.g., temperature, cloud cover, wind) degrade entanglement in
volatile ways.  
Our protocol queries a pre-built model of link
degradation to try to predict how the link will behave, and then sets
the controllable parameters to compensate for the expected
degradation.

Our main result is a classical control protocol that boosts the practical utility of wide-area quantum links by maximizing the rate of entanglement distribution at a consistent, user-defined fidelity.
The protocol requires no additional quantum hardware and introduces minimal classical overhead.
To make things concrete, we analyze an instance of our protocol based around a 64-km deployed fiber in the Washington DC metropolitan quantum network (DC-QNet)~\cite{mckenzie_sync-dc-qnet_2024}.
Our link layer protocol directly benefits wide-area quantum networks based on quantum repeaters, as these networks operate by swapping elementary entanglements and therefore the rate and fidelity of each physical hop directly constrains end-to-end performance.

The rest of this paper is structured as follows: we introduce relevant hardware
properties and discuss the specific quantum hardware used in our
experiments in Section~\ref{en-dist}.  In Section~\ref{exp-hw}, we present modeling of the components used by our
adaptive protocol.  Section~\ref{sec:opt} introduces a sequence of
optimization problems that decompose different aspects of entanglement
distribution, while Section~\ref{sec:protocol}
describes the control protocol itself.
Section~\ref{results} evaluates the protocol against existing techniques, Section~\ref{sec:discussion} discusses generalization to other hardware and higher network layers, and Section~\ref{conc} concludes.

\section{Entanglement Distribution at the Physical Layer} \label{en-dist}

The elementary unit of a quantum network is a physical link that generates and transmits entanglement.
The hardware involved in this point-to-point entanglement distribution consists of a source, link, and receiver, and is directly relevant for applications like entanglement-based quantum key distribution~\cite{rusca_2024_cryptography}, sensing, and metrology~\cite{Zhang_2021}, wherein the receiver immediately characterizes entangled states. This framework can be extended to more sophisticated link models, like those in future quantum repeater networks, which rely on additional technologies like quantum memories~\cite{wei_review_2022}.

\subsection{Entanglement Sources} \label{sec:sources}

An \emph{entanglement source} is a device capable of generating pairs of entangled qubits. 
For long-distance quantum communication, information is encoded into one of many physical degrees of freedom within individual photons.
These ``flying qubits’’ must be generated with dedicated quantum hardware.

Entanglement can be generated via clever manipulation of individual qubits from \emph{single-photon sources} (as in~\cite{dlcz_2001}), or from sources that natively produce entangled photon pairs.
In practice, entanglement generation is typically a probabilistic process, though technology for on-demand, deterministic sources is improving~\cite{eisaman_review_2011,li_review_2023}.
Sources can be characterized in part by their photon generation rate, single-photon purity (quantifying the degree of errant multi-photon emission), and the resulting quantum state fidelity~\cite{bienfang_single-photon_2025}.

Many source properties are fixed during the engineering process; however, these systems still typically allow some degree of tunability.
In this work, we consider a simple example of modulating the pump laser power for a source based on spontaneous parametric down-conversion (SPDC); details are covered in Section~\ref{sec:sd-model}. 
Sources based on four-wave mixing or emission from solid state quantum memories also often exhibit tunability via pump laser power.
Additional examples of controllable parameters include central wavelength and bandwidth temperature tuning for both SPDC and quantum dot sources.

\subsection{Quantum Receivers} \label{sec:receivers}

Once an entangled qubit reaches its destination, the information may be stored, measured, or processed through some logical operation.
Ubiquitous across all receiver operations is the \emph{single-photon detector}, a device capable of identifying the presence of individual photons~\cite{bienfang_single-photon_2025}.

Single-photon detectors are characterized by their efficiency and dark count rate (among other measures) which indicate the true-positive and false-positive rate of photon detection, respectively.
State-of-the-art performance is achieved using superconducting nanowire single-photon detectors (SNSPDs), reaching upwards of 98\% efficiency at standard telecom wavelengths~\cite{reddy_2020_snspd,li_surpassing_2025}.

Similar to sources, 
detectors can offer some degree of tunability. For example, SNSPD bias current can be adjusted to achieve (limited) increased detector efficiency. This comes at the expense of increased dark counts, but may still yield an overall advantage in some applications~\cite{nunn_2021_heralding}.
Additional tradeoffs can be introduced in the processing of photodetection signals, such as tuning the ``coincidence window'' to qualify the simultaneity of detection events~\cite{coincidence-window}, or the duration of a measurement for state characterization, trading integration time for reduced statistical uncertainty.

\subsection{Polarization Drift Across a Link} \label{sec:polarization}

Ideal quantum channels would perfectly preserve quantum states between source and receiver, but real-world links invariably introduce some degree of loss and decoherence of quantum information as photons interact with the environment.

A central characteristic of optical communication, classical or quantum, is \emph{polarization}.
Physically, polarization describes the direction of oscillating electromagnetic fields that comprise light. For most quantum networking purposes, maintaining consistent polarization between all parties is strictly required, especially when the polarization itself encodes the quantum information.
This poses a great challenge for transmission over realistic quantum links, whereby polarization state fidelities inevitably decay without intervention.

The objective of polarization stabilization is to ensure a photon's state of polarization (SOP) remains unchanged from source to receiver.
For our purposes, the SOP can be described by three parameters $(S_1, S_2, S_3)$ which define a unit vector known as a \emph{Stokes vector}.
The received vector $\Vec{v}'=(S_1',S_2',S_3')$ will be altered by optical fiber imperfections and environmental conditions. These can also change over time --- a phenomenon known as \emph{polarization drift}. While the magnitude of $\Vec{v}'$ may change\footnote{via polarization-dependent loss or polarization mode dispersion~\cite{gisin_pmd-pdl_1997}.}, the most significant and dynamic changes are usually in the vector direction. The angle between the desired and received vectors is the drift angle $\theta$, from which the polarization fidelity $\Ftrans$ is calculated~\cite{jozsa_1993}:
\begin{equation} \label{eqn:angle2fid}
    \Ftrans = \frac12\left(1 + \cos(\theta)\right)
\end{equation}
The exact drift conditions are unique to each link, and can be highly variable over time depending on the environment.

Significant effort goes into maintaining polarization over a quantum link.
This process commonly works by sending classical reference signals, measuring the change in polarization of those signals, then applying a correction to offset this change~\cite{fieldqkd_2018_li,kucera_demonstration_2024}.
In the following subsection, we discuss a commercial solution to polarization stabilization that operates on this principle.

\subsection{Concrete Hardware Instance}

\begin{figure*}[t]
    \includegraphics[width=\linewidth]{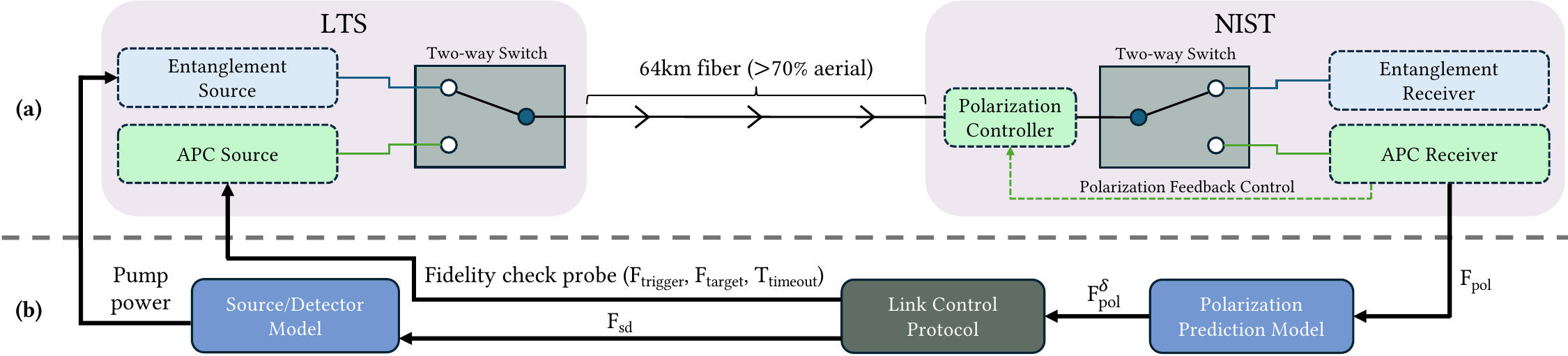}
    \caption{(a) Schematic of our quantum link. Entanglement generation and automated polarization compensation (APC) signals are time multiplexed. Measurements at the APC receiver inform corrections at the polarization controller.
    (b) Feedback loop used for link control. Polarization fidelity ($\Ftrans$) measured by the APC acts as input to the link control protocol and its associated models, which in turn determine hardware parameters at the source site.
    \label{fig:experimental-setup}}
\end{figure*}

\begin{figure}[t]
    \includegraphics[width=.9\linewidth]{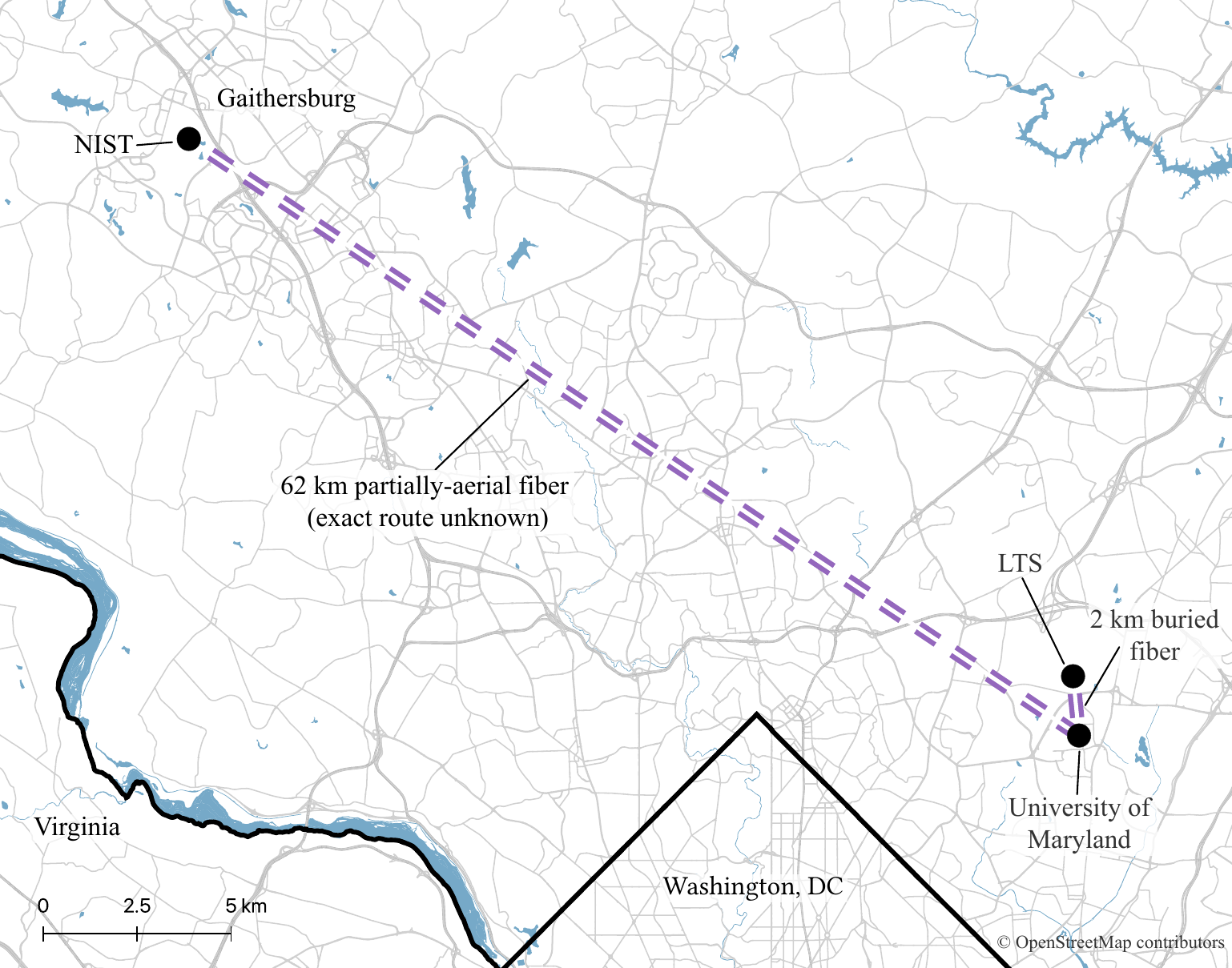}
    \caption{Metropolitan context of the 64km deployed fiber link, part of the Washington DC metropolitan quantum network (DC-QNet). The source is located at the Laboratory for Telecommunication Sciences (LTS) in College Park, MD, USA. The route passes through a switch at the University of Maryland, then continues through a 62km partially-aerial fiber to the National Institute of Standards and Technology (NIST) in Gaithersburg, MD, USA. The exact routes of these two fibers are not known and therefore sketched as straight dashed lines.
    \label{fig:link-map}}
\end{figure}

To make the ideas of this paper concrete, we analyze a protocol instance based around hardware available to us, described here and sketched in Figure~\ref{fig:experimental-setup}.
However, the same ideas extend naturally to many other experimental setups, see Sections~\ref{sec:generality} and~\ref{sec:discussion-generality}.

Our simulations are driven by polarization data gathered over a 64km quantum link in the Washington DC metropolitan quantum network (DC-QNet)~\cite{mckenzie_sync-dc-qnet_2024}.
The link consists of a 2km buried optical fiber connecting the Laboratory for Telecommunication Sciences (LTS) to a switch at the University of Maryland, which routes photons through a 62km partially-aerial fiber to the National Institute of Standards and Technology (NIST).
Figure~\ref{fig:link-map} sketches the route of this link.
Transmission loss over this fiber link was measured to be approximately 18dB.
Over 70\% of the fiber is aerial, which leads to greater polarization instability compared to buried fiber due to its more direct interaction with environmental changes.

A schematic of our entanglement source and detector setup is shown in Figure~\ref{fig:source-diagram-simplified}.
Our source of polarization-entangled photon pairs is based on SPDC, the most mature (and therefore widely-deployed) class of sources used for quantum communications~\cite{eisaman_review_2011}.
The power of the input \textit{pump laser} driving the source can be easily modulated to determine the rate and fidelity of entanglement generation (see Section~\ref{sec:sd-model} and Figure~\ref{fig:sd-characterization}).
A more detailed description and diagram can be found in Appendix~\ref{sec:source-detailed}.

Our receiver is comprised of a pair of polarization analyzers and SNSPDs (single-photon detectors). The analyzers allow us to perform full quantum state characterization (tomography) on the polarization-encoded photons emitted from the source~\cite{tomography_2005}. From this, we compute the entanglement fidelity to an ideal Bell state. Note that the procedure in Figure~\ref{fig:source-diagram-simplified} depicts local characterization of the source-receiver system. For entanglement distribution, one qubit is detected locally near the source, while the other is instead measured at a remote location across a fiber link (see Figure~\ref{fig:experimental-setup}a).

\subsubsection{Polarization Drift}

In our experiments, polarization drift causes the fidelity $\Ftrans$
of our polarization-encoded qubits to degrade over time.  Our protocol
relies on a sufficient understanding of the \textit{rate of change} in
polarization (the ``drift rate'') in order to make informed parameter
adjustment decisions.  Figure~\ref{fig:polarization} displays measured
polarization drift over a 68 hour period over the link.  The plot
shows the raw polarization Stokes parameters during this period with
the distilled mean drift rate $\lvert\frac{d\theta}{dt}\rvert$ overlaid.  Drift rates are strongly correlated with time of
day,
with significantly higher drift over daytime hours.

\subsubsection{Polarization Compensator}

During operation, the polarization drift on our link is measured and corrected using a commercial\footnote{Commercial equipment and software referred to in this work is identified for informational purposes only, and does not imply recommendation of or endorsement by the National Institute of Standards and Technology, nor does it imply that the products so identified are necessarily the best available for the purpose.} Automated Polarization Compensation (APC) device~\cite{qunnect-apc}. 
The APC measures polarization using in-band classical signals, meaning that no quantum data can be transmitted while it is in operation.
Figure~\ref{fig:experimental-setup}a illustrates how the path must be fully switched away from the entanglement source and detector while the APC is in use.

The APC may be probed at any time.
On each probing, three parameters must be passed to the device: (1) a triggering fidelity threshold $\Ftrigger$, (2) a target fidelity threshold $\Ftarget$, and (3) a timeout duration $\timeout$.
When the APC is probed, it initiates a ``fidelity check''.
A fidelity check takes a constant amount of time, 44ms for our link, and measures the current polarization fidelity $\Ftrans$.
If $\Ftrans > \Ftrigger$, the APC concludes operation and entanglement distribution resumes.
However, if $\Ftrans \leq \Ftrigger$, the device initiates a compensation routine, during which it adjusts the polarization controller in an attempt to raise the polarization fidelity until $\Ftrans \geq \Ftarget$. This routine takes an unspecified amount of time and is not guaranteed to succeed.
If the target fidelity has not been met after $\timeout$ seconds, the routine aborts and entanglement distribution resumes.

\begin{figure}
    \centering
    \includegraphics[width=.75\linewidth]{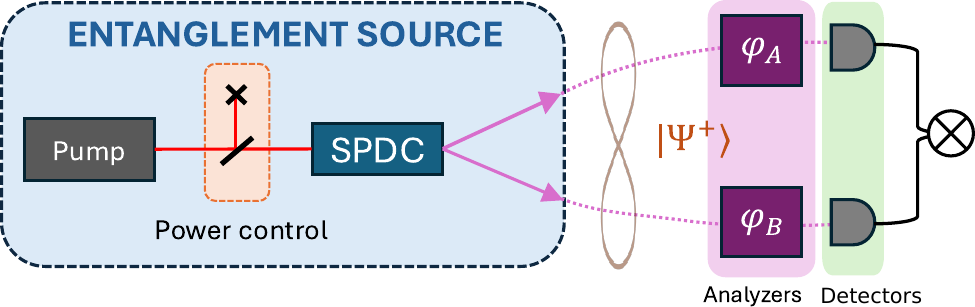}
    \caption{Schematic of experimental setup used to characterize SPDC entanglement source and receiver with controllable pump laser power. A complete diagram and its description can be found in Appendix~\ref{sec:source-detailed}.}
    \label{fig:source-diagram-simplified}
\end{figure}

\subsection{Generality of the Control Abstraction} \label{sec:generality}

While the results in this paper are evaluated using a particular wide-area link with a particular source, detector, and polarization controller, the control logic is defined over a hardware-agnostic link interface.
This interface consists of three replaceable components: (1) a set of controllable hardware settings that define a tradeoff between entanglement generation rate and fidelity, (2) a model for time-varying link degradation, and (3) a compensation mechanism and corresponding model mapping corrective actions to expected fidelity improvement and overhead.

\begin{figure}
    \begin{subfigure}{.48\columnwidth}
        \includegraphics[width=\columnwidth]{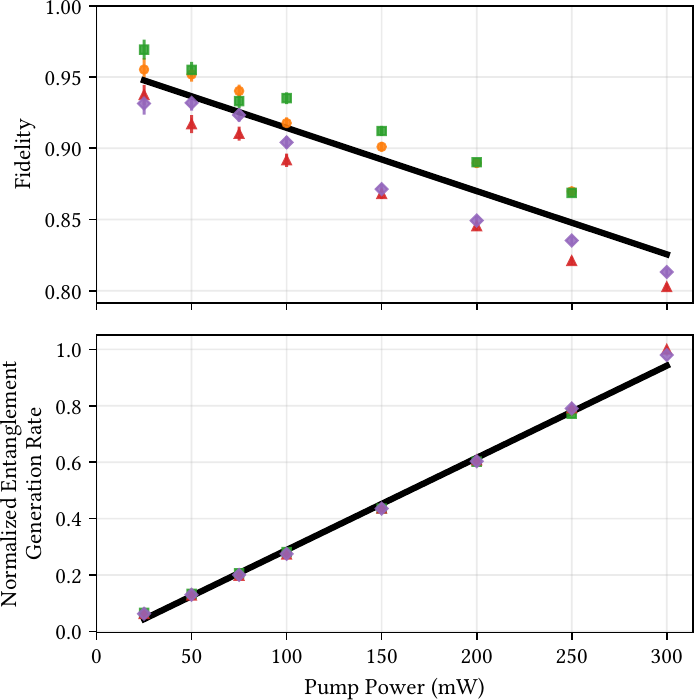}
        \caption{Characterization of an in-house entanglement source based on spontaneous parametric down-conversion (SPDC).
        \label{fig:sd-characterization}}
    \end{subfigure}
    \hfill
    \begin{subfigure}{.48\columnwidth}
        \includegraphics[width=\columnwidth]{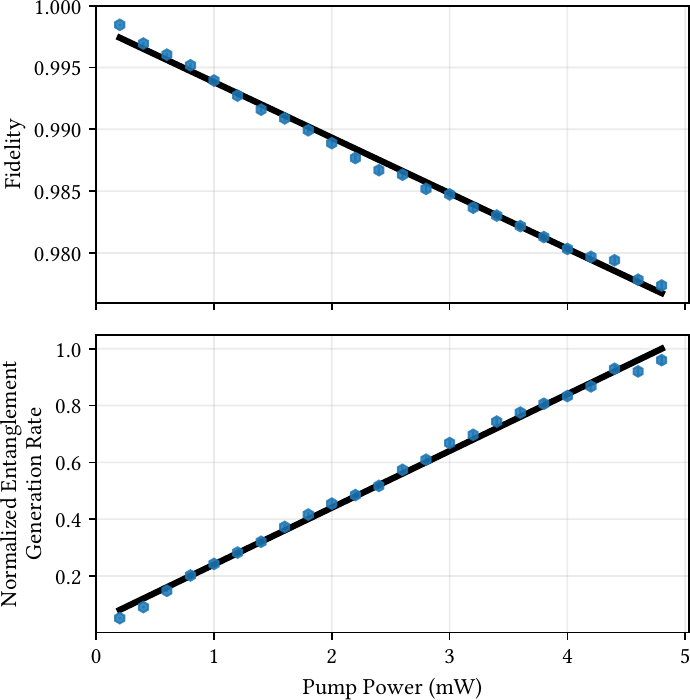}
        \caption{Characterization of a commercial Qunnect Qu-SRC entanglement source~\cite{qu-src} based on four-wave mixing.}
        \label{fig:sd-characterization-qusrc}
    \end{subfigure}
    \caption{Controllable rate-fidelity tradeoffs across two entanglement source technologies. In both sources, increasing pump power increases entanglement generation rate but reduces entanglement fidelity. The simulations in this paper use the SPDC characterization in (a).}
    \label{fig:tradeoffs}
\end{figure}

For the source-detector setup, the relevant requirement is a controllable rate-fidelity operating frontier.
In our evaluated system, pump power provides this control.
Figure~\ref{fig:tradeoffs} shows this behavior for our SPDC source and for a second commercial source based on four-wave mixing~\cite{qu-src}.
Four-wave mixing probabilistically generates entangled photon pairs by a fundamentally different physical mechanism than SPDC~\cite{eisaman_review_2011}, yet importantly, both systems exemplify the same control abstraction: 
increasing pump power increases the entanglement generation rate while reducing entanglement fidelity over the operating regime.

Note, in addition, that our approach extends well beyond these examples.
The same abstraction encompasses many quantum memory-based links as well, with controllable rate-fidelity tradeoffs common in links that entangle pairs of trapped ions~\cite{trapped-ion-remote-entanglement}, neutral atoms~\cite{neutral-atom-remote-entanglment}, and color centers in diamond~\cite{link-layer-protocol, long-baseline-interferometry}. 
In all these examples, the tradeoffs are modulated by source laser power;
however, this need not be the case, as many other control knobs exist depending on source and detector configurations. We list several such parameters in Sections~\ref{sec:sources} and~\ref{sec:receivers}.
The protocol also does not require the rate-fidelity relationship to be linear or one-dimensional, as the same role can be played by any calibrated set of operating points, including a multidimensional Pareto front over rate and fidelity.

On the link side, the relevant requirement is time-varying degradation that can be measured or conservatively bounded over the control horizon, together with a corrective action that can improve the link state at some cost.
Polarization drift is a ubiquitous example of a link-degrading phenomenon in optical quantum communication, and its magnitude depends on link-specific factors such as length and deployment environment.
These factors enter the protocol through the learned drift model rather than through hardware-specific assumptions,
so the same approach applies to other environments, e.g., buried or undersea fibers.
A different deployment would replace or update the training data and prediction model, but the structure of the controller remains unchanged.

Similarly, our use of a commercial APC device is one instance of a more general compensation primitive.
The protocol’s only requirement is a way to estimate when compensation is worth its cost through a sufficiently validated model.
Other compensation hardware may expose different parameters or incur different overheads, but these differences can be incorporated by reworking the compensation response model.
The APC model used in our simulator is therefore an instantiation of this interface, not a protocol requirement.

\section{Modeling the Evaluated Quantum Link} \label{exp-hw}

We now instantiate the three required interfaces described in Section~\ref{sec:generality} for the deployed link used in our evaluation
by describing the modeling of relevant hardware components.
These models abstract the performance characteristics of each component and
are used as input to our control protocol described in Section~\ref{sec:protocol}.

\subsection{Source-Detector Model} \label{sec:sd-model}

For the evaluated deployment, we instantiate the source-detector model using empirical data from an SPDC source designed to produce polarization-entangled photon pairs. In Figure~\ref{fig:sd-characterization}, our source is characterized for a range of input pump powers, from 25 to 300 mW. The upper subplot shows the quantum state fidelity to an ideal Bell state, computed from the tomographic measurement shown in Figure~\ref{fig:source-diagram-simplified}. The lower subplot shows the rate of coincident detection events at each of the two SNSPDs, normalized to the maximum observed value. This rate depends on the efficiency of the SPDC process, coupling into optical fiber, detector efficiency, and other link losses between source and receiver. Details of the data in Figure~\ref{fig:sd-characterization} 
and its processing are given in Appendix~\ref{sec:sd-model-detailed}.

As pump power increases, the photon pair generation rate increases linearly in this regime. However, due to the probabilistic nature of SPDC, increasing the probability of emission per each laser pulse also increases \emph{multipair} emission-- the likelihood that more than one photon pair is generated from a single pulse. This effectively adds noise and reduces the entanglement fidelity $\Fsource$, as the measured polarization correlations are masked by independently generated pairs that are not correlated to one another. The maximum fidelity in the source-detector setup of Figure~\ref{fig:source-diagram-simplified} (average $\Fsource\approx0.95$) is observed at low pump powers where multipair probability is negligible, but the rate is also reduced. Hence, the pump power serves as a control parameter to appropriately balance rate and source fidelity.

\subsection{Polarization Drift Prediction Model} \label{sec:drift-model}

\begin{figure*}[t]
    \centering
    \includegraphics[width=\textwidth]{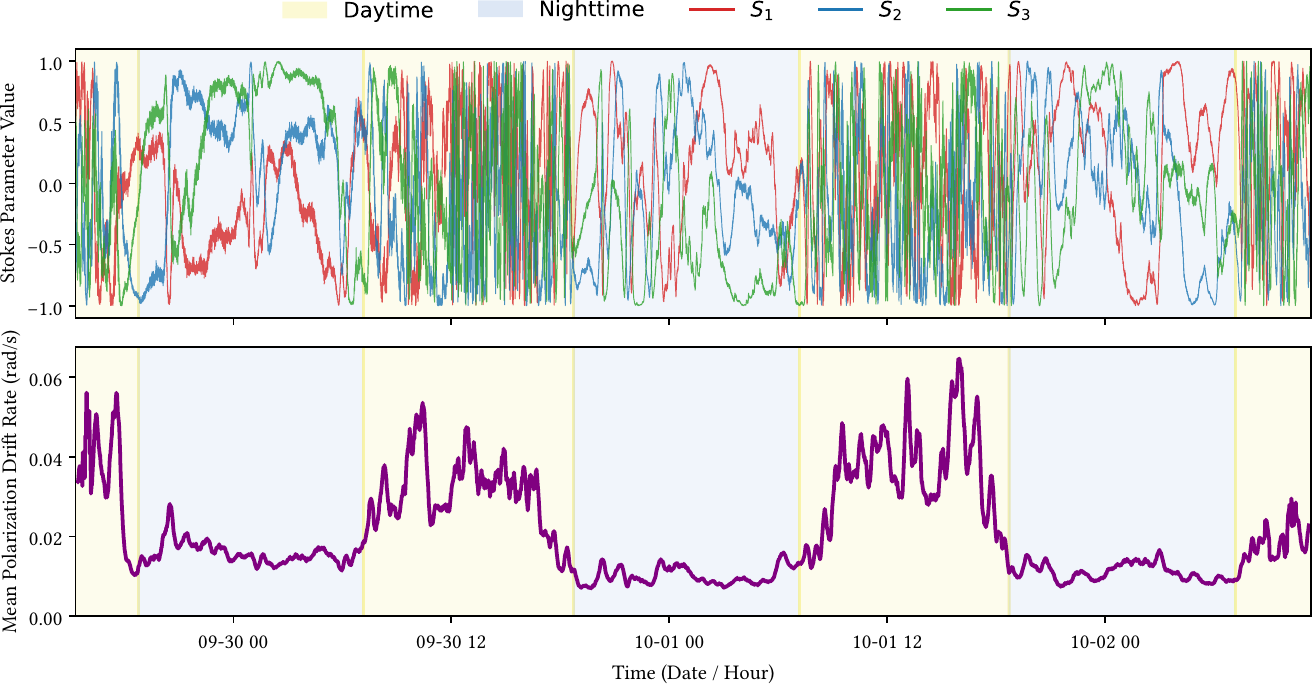}
    \caption{Measured polarization drift over 68 hours on a 64km fiber link. The top figure shows the evolution of raw Stokes parameters over time; the bottom plot shows the calculated drift rate averaged over 15 minute intervals. On average, drift is significantly higher during the day.
    }
    \label{fig:polarization}
\end{figure*}

Polarization drift is a complex physical process that we do not attempt to model precisely in this paper, although related models have been constructed~\cite{banner2025bifrost}.
In principle, one could model polarization drift by monitoring the relevant environmental factors
(wind, incident sunlight, vibrations from vehicles, etc.); however, in practice, it is impractical over a link with 10s of kilometers of aerial fiber.
Since we are only interested in the aggregate change in polarization over time, a simpler empirical method suffices for our purposes.

To develop our model, we instead turn to a more readily accessible variable: measured drift rates.
We find that polarization drift over a short period of time is strongly correlated with drift over a subsequent short period of time.
We build our model empirically from a dedicated experiment which tracked the state of polarization over a period of 68 hours\footnote{Given a sufficiently long operation time, such a model could be built without a dedicated experiment by gathering drift statistics during operation.},
shown in Figure~\ref{fig:polarization}.
The model is trained on 24 of these hours; we analyze the sensitivity to training dataset size in Appendix~\ref{sec:appendix-drift-model}.

The training of our model is simple: for a pair of durations $(\Delta t_1, \Delta t_2)$, we sample a time $t$ and find the corresponding three stokes vectors $v_{t-\Delta t_1}, v_t, v_{t+\Delta t_2}$.
Then we compute the two consecutive drift angles as $\theta_1 = \texttt{angle\_between}(v_{t-\Delta t_1}, v_t)$ and $\theta_2 = \texttt{angle\_between}(v_t, v_{t+\Delta t_2})$.
Repeating this for $N=10^6$ randomly chosen values of $t$ results in a distribution of drift angles $\theta_2$ given parameters $(\Delta t_1,\Delta t_2,\theta_1)$.
We repeat this for 64 representative values of $(\Delta t_1, \Delta t_2)$ to create a general and practical model.

\subsection{Modeling the Polarization Compensator} \label{sec:apc-model}

\begin{algorithm}[b]
\caption{Simulation of Polarization Compensation}
\label{alg:apc-sim-alg}
\begin{algorithmic}[1]
    \Require the initial fidelity $F_{0,\text{pol}}$, the target fidelity $\Ftarget$, the start time $t_0$
    \Ensure total compensation time, final fidelity

    \State $\theta, \theta_\text{target} \gets \texttt{angles\_from\_fids}(F_{0,\text{pol}}, \Ftarget)$
    \Comment{Convert fidelities to drift angles via Eq.~\ref{eqn:angle2fid}}
    \State $t \gets t_0$

    \While{$\theta>\theta_{\text{target}}$}
        \State $\theta = \theta - \frac\eta2\sin\theta + \Delta \cdot d(t)$ 
        \Comment{Simulate a noisy gradient descent step}
        \State $t \gets t+\Delta$
    \EndWhile

    \State \Return $t-t_0, \texttt{fid\_from\_angle}(\theta)$
\end{algorithmic} 
\end{algorithm}

Simulating the effects of the APC requires a model to predict its run time and resulting fidelity upon termination, given the current link state.
These aspects of the device have not been studied, so we developed a model based on conversations with the device manufacturer and 70 hours of experimental data from an APC test over a similar link (62km aerial fiber).

The details of the APC's compensation algorithm are proprietary, so we make several assumptions in the construction of our model.
We know that the routine is gradient-descent-based, repeatedly measuring the polarization state and applying corrections until the fidelity target is met or the routine times out.
We assume vanilla Riemannian gradient descent, which is designed for optimization on smooth manifolds such as the surface of the 3-dimensional unit sphere of Stokes vectors.
The update step of this algorithm is 
\begin{equation}
    \theta_{k+1} = \theta_k - \frac\eta2\sin\theta_k
\end{equation}
where $\theta_k$ is the angle between the ideal and realized vectors at iteration $k$ of gradient descent and $\eta$ is the step size.

Second, we assume that the effect of polarization drift on compensation can be modeled by a noise term within each gradient descent step.
We assume this term is equal to the product of the current drift rate $d(t)$ and the duration of each step $\Delta=27.7$ms.

The modeled compensation routine is shown in Algorithm~\ref{alg:apc-sim-alg};
more details on the fit to experimental data can be found in Appendix~\ref{sec:appendix-apc-hist}.
We note that the results in this paper assume this model, and deviations from this model may affect the results upon deployment.

\newcommand{\opt}[1]{{\em Opt-{#1}}\/\xspace}

\section{Optimizing Entanglement Distribution}\label{sec:opt}

Given an understanding of the available hardware and modeling techniques, described in the previous section for one deployed link, we now introduce a more general framework that decomposes entanglement generation into a sequence of optimization problems.
In this section, we focus on optimizing the performance of a single link.
We extend the framework to consider entanglement across multiple links in Section~\ref{sec:discussion-opt3}.

\subsection{Opt-0: Optimizing Controllable Parameters} 

We first consider a stable experimental setting in which link drift is
negligible or ignored, such as a short, closed laboratory link or a
brief experiment with pre- and post-calibration checks.  Even in this
regime, source- and receiver-side parameters must be tuned to maximize
entanglement rates under a minimum (and possibly time-varying)
fidelity constraint.  We define \opt{0} as the problem of selecting
controllable parameters (e.g., pump power at the source, integration
time at the receiver) to maximize entanglement rate subject to a
target fidelity.

An \opt{0}-type optimization has been integrated into a quantum link
layer protocol~\cite{link-layer-protocol,
  experimental-link-layer-protocol}, where a parameterized microwave
pulse at the source is used to control the rate-fidelity tradeoff.
The class of problems posed by \opt{0} is more general, as it
encompasses multiple controllable parameters at both the source and
receiver and potentially more complex tradeoffs.  
Nevertheless, these remain the primary studies to consider
link-level control in this context.

\subsection{Opt-1:  Managing Uncontrollable Parameters}

For wide-area entanglement distribution, links are inevitably subject
to time-varying and unpredictable noise, particularly for above-ground
deployments exposed to environmental factors such as temperature,
wind, and human activity.  Some effects exhibit predictable patterns
(e.g., diurnal temperature cycles), while others do not, complicating
long-duration experiments.

We define \opt{1} as the problem of optimizing the mitigation of these
uncontrollable, performance-degrading link parameters.  Prior work has
addressed physical-layer aspects of \opt{1} over long-distance links,
including clock synchronization~\cite{mckenzie_sync-dc-qnet_2024}
through the IEEE-1588 Standard (PTP~\cite{ptp} and related White
Rabbit~\cite{white-rabbit} protocols) and polarization stabilization
using commercial compensation
devices~\cite{yicheng-entanglement-distribution,qunnect-apc}.
However, prior to this work, these techniques have not been
systematically optimized, particularly with respect to their impact on
end-to-end system performance.

Solutions to \opt{1} must maintain acceptable operating conditions
while balancing downtime, compensation accuracy, and interference with
the quantum signal.  Compensation often requires pausing quantum data
transmission, making its frequency and duration critical design
parameters.  Thus, \opt{1} involves trading off improved link quality
against compensation overhead to maximize overall link utility.

\subsection{Opt-2: Joint Optimization}

Finally, consider again a general wide-area quantum link.  Practical
entanglement distribution can only be achieved if the link is able to
consistently produce high-fidelity entanglement at a sufficient rate.
Achieving this requires tuning of controllable parameters according to
application-specified fidelity requirements and the current state of
the link.  Further, invocation of any link compensation algorithm must
take into account the overhead of drift correction in conjunction with
the current link state.  This is problem \opt{2}: joint optimization
of controllable parameters while compensating for the uncontrollable.
A salient feature of \opt{2} solutions is the feedback loop between
compensation overhead and controllable parameter settings, as both
must be considered in concert to achieve high entanglement rates.

The focus of this paper is a protocol for \opt{2}.  Part of our work
concerned with source control is similar
to~\cite{link-layer-protocol}, though the hardware involved, as we
have described, is markedly different.  The primary focus of our
protocol is to explore the feedback loop intrinsic in solutions for
\opt{2}: scheduling compensations based on link performance, which
depends on the parameter settings and compensation quality.  We
describe this protocol next.

\section{Dynamic Control Protocol} \label{sec:protocol}

The experimental setup described above defines a control problem.
Our link protocol therefore has access to five control variables:
\begin{itemize}
    \item Source-detector parameter: (1) pump laser power
    \item Drift parameters: (2) timing of fidelity checks, (3) APC triggering fidelity threshold, (4) APC target fidelity threshold, (5) APC timeout duration
\end{itemize}

At time $t$, the link distributes entanglement at a rate of $r(t)$ entangled pairs per second with approximate\footnote{The true end-to-end fidelity would be estimated experimentally by a prolonged measurement of many copies of the transmitted quantum states, destroying the entanglement~\cite{photonic-state-tomography, no-cloning}. In contrast, this $F(t)$ model is efficiently computable for our real-time protocol, and should well approximate the true fidelity, as shown in Appendix~\ref{sec:appendix-fidelity}.}
fidelity $F(t)\approx\Fsource(t)\Ftrans(t)$. 
The objective of a protocol here is to maximize the expected long-run average entanglement distribution rate subject to a hard minimum fidelity constraint $\Fmin$.
Entangled pairs are discarded when the fidelity requirement is not met; i.e., $r(t)=0$ whenever $F(t)<\Fmin$\footnote{Discarding below-threshold pairs is an evaluation convention; in practice, the link cannot know when a pair is below threshold, as any attempt to measure its fidelity would cause the information to collapse. This further motivates our desire to stay above threshold.}.
The choice of this strict fidelity requirement is to ensure robust entanglement generation at the link level, as applications of quantum networks require entanglement of sufficient fidelity~\cite{link-layer-protocol} and protocols involving multiple links typically assume a constant fidelity for each link.

\paragraph{Static Protocols}

Existing entanglement distribution protocols of this type make
straightforward use of available resources and APIs by setting
constant values for the five control variables. A recent experiment
demonstrating state-of-the-art entanglement distribution across a
metropolitan scale link uses constant values for all five control
variables~\cite{yicheng-entanglement-distribution}.

\begin{table}[]
\centering
\caption{Summary of variables used in this paper.}
\begin{tabular}{|l|l|}
\hline
\textbf{Variable} & \textbf{Definition} \\ \hline
$F$  & end-to-end fidelity      \\ 
$r$  & end-to-end entanglement distribution rate      \\ 
$\Fmin$  & minimum fidelity requirement    \\ 
\hdashline
$\Fsource$  & source-detector fidelity     \\ 
$\Ftrans$  & polarization fidelity      \\ 
$\Ftrans^\delta$  & $\delta$-conservative prediction of $\Ftrans$     \\ 
\hdashline
$\Ftrigger$  & APC triggering fidelity threshold      \\ 
$\Ftarget$  & APC target fidelity threshold      \\ 
$\timeout$  & APC timeout duration      \\ 
\hdashline
$\Teff$  & expected time until the next fidelity check      \\ 
$\Bar{r}$  & average rate within a period      \\ 

\hline
\end{tabular}

\label{vars}
\end{table}

\subsection{Intuition}

It is useful to think of the system as two independent components:
\begin{enumerate}
    \item The \textbf{source and detector} form a single logical unit responsible for the creation and measurement of Bell states. The rate $r(t)$ and fidelity $\Fsource$ of entangled pair generation are fully and continuously controllable by the pump laser power (see Section~\ref{sec:sd-model}, Figure~\ref{fig:sd-characterization}).
    \item The \textbf{transmission} medium responsible for distributing entangled photons between the source and detector is not directly controllable and only adds noise (and loss) to the entanglement generated through subsystem (1). Much of the effect of transmission is determined by the environment, can only be measured through active probing, and can only be controlled through active compensation (see Figure~\ref{fig:polarization}).
    
\end{enumerate}

From this decomposition, it becomes clear that the role of the pump power should be to counteract the effect of polarization drift.
Since $\Ftrans$ tends to decrease over time, we want to lower the pump power over time in order to raise $\Fsource$ and therefore sustain $F(t)=\Fsource(t)\Ftrans(t)\geq\Fmin$.
This approach will decrease our end-to-end rate over time, as a consequence of the decreasing pump laser power.

Ideally, we would like to keep our pump power as high as possible to
maximize the end-to-end rate.  However, this is likely sub-optimal
as the polarization drift is unknown.
Instead, we use our polarization drift prediction model
to estimate the extent of polarization drift, and select pump power
conservatively to ensure that we maintain $F(t)\geq\Fmin$ with high
probability.  Our methodology for meeting this criterion is explained
in Section~\ref{sec:power-policy}.

This conservative power adjustment, while necessary to avoid
high-penalty $\Fmin$ violations, tends to reduce our rate more than is
strictly necessary.  In addition, the uncertainty in our polarization
estimate increases with time.  In order to correct possible
sub-optimality, it is desirable to invoke a fidelity check to learn
the current value of $\Ftrans(t)$.  Doing so allows us to reset the
pump power to the optimal value and gives us updated information on
the current polarization drift levels.  However, we do not want to
invoke this capability too often, as we want to balance the
information gain of a fidelity check with the downtime it incurs.  The
full policy is explained in Section~\ref{sec:fid-check-policy}.

After some amount of time, we should perform polarization compensation
in order to mitigate the effects of polarization drift, which forces
our rate to generally decrease over time.  To do so, we set the APC
triggering fidelity such that we trigger a compensation when the
fidelity has drifted low enough that the increase in rate from
compensating outweighs the opportunity cost of downtime during
compensation.  We also set the value of the target fidelity threshold
and timeout duration to balance the increase in $\Ftrans$ with
downtime.  The details of our policy for setting the APC device
parameters are given in Sections~\ref{sec:timeout-policy} and
\ref{sec:triggering-policy}.

The control techniques here are purely classical and are partially analogous to some adaptive resource control protocols in classical networking. We discuss these comparisons in Appendix~\ref{sec:appendix-classical-comparison}.

\subsection{Pump Power Control Policy} \label{sec:power-policy}

Given perfect information of $\Ftrans(t)$, the optimal pump power control policy is to choose the pump power such that $F(t)=\Fmin$, i.e.,
\begin{equation} \label{eqn:Fsd-calculation}
    \Fsource = \Fmin / \Ftrans.
\end{equation}
This choice is guaranteed to maximize rate subject to the minimum fidelity constraint.

During execution, however, the exact value of $\Ftrans$ is only known immediately after a fidelity check.
Given this uncertainty, our polarization drift model allows us to maintain a belief distribution over $\Ftrans(t)$.
Using this, we control the pump power conservatively such that
\begin{equation} \label{eqn:conservative-pump-power}
    \Pr[F(t)<\Fmin]\leq\delta
\end{equation}
for a fixed confidence parameter $\delta\in(0,1)$.
Equivalently, power is chosen as if the transmission fidelity were equal to the $\delta$-quantile of the belief distribution.
In our simulations, we select $\delta=0.10$, meaning that we assume a polarization drift rate in the $90^{\text{th}}$ percentile according to our model.

\subsection{Fidelity Check Initiation Policy} \label{sec:fid-check-policy}

A fidelity check returns the exact current value of $\Ftrans$ after a fixed downtime duration $t_{\text{check}}$, during which the rate $r(t)=0$.
Fidelity checks may be initiated at any time.

The protocol schedules fidelity checks adaptively, based on a value-of-information criterion.
Let $r(\Ftrans^\delta)$ denote the rate achieved using our current predicted $\delta$-conservative polarization fidelity.
Let $\Teff$ denote the expected time until the next check, and
let $\Lambda_1$ and $\Lambda_2$ be the expected throughput over the next $\Teff$ seconds if we do and do not perform a check, respectively.
A fidelity check is initiated when
\begin{equation} \label{eqn:fid-check-condition}
    \Lambda_1 - \Lambda_2 \geq r(\Ftrans^\delta)t_{\text{check}}.
\end{equation}

This condition compares the expected rate gain from resolving uncertainty to the opportunity cost of downtime.
Fidelity checks are therefore performed only when uncertainty induces a sufficiently large rate penalty.
$\Lambda_1,\Lambda_2$ may be calculated through estimates of future rates using the polarization drift predictor --- see Appendix~\ref{sec:appendix-rate-integration} for details on efficient computation of $\Lambda_i$.

\subsection{Fixing Compensation Duration} \label{sec:timeout-policy}

Our choice of polarization compensation target fidelity $\Ftarget$ and timeout duration $\timeout$ is based on a simple observation: that setting 
\begin{equation}
    \Ftarget = 1.0 ~~;~~\timeout = c
\end{equation}
for some constant $c$ ensures that the compensation routine will take time $c$.
To see this, note that the APC's gradient descent approach combined with continuous drift effectively ensures $\Ftrans(t)<1.0$ at all times $t$.
In our simulations, we choose $c:=1$ second.

This choice is unconventional but carries several benefits.
First, assuming a constant compensation duration greatly simplifies the analysis of optimal compensation initiation, allowing for a much cleaner triggering policy (see Section~\ref{sec:triggering-policy}).
It also removes the need for potentially complex control policies governing $\Ftarget$ and $\timeout$.

Second, limiting compensation time makes efficient use of the diminishing returns seen during the compensation routine.
From Section~\ref{sec:apc-model}, we expect that the increase in fidelity decays exponentially with each gradient descent iteration, while the loss of throughput during compensation is linear in time.
As a result, the net benefit of additional compensation time is strictly decreasing; fixing the compensation time truncates this low-value tail in a principled manner.

Finally, this policy is robust against unknown, time-varying polarization drift.
From Section~\ref{sec:apc-model}, we find that a high drift rate can prevent the compensator from reaching the target fidelity; this phenomenon commonly results in expensive compensation routines trying to reach an unachievable $\Ftarget$ (see~\cite{yicheng-entanglement-distribution}).
In contrast, a fixed-duration policy avoids implicit assumptions about the achievable polarization fidelity and guarantees bounded downtime regardless of drift conditions.

In Appendix~\ref{sec:appendix-dynamic-compensation}, we consider an algorithm that dynamically selects compensation times and find that such an approach adds substantial complexity for only marginal benefit.

\subsection{Polarization Compensation Initiation} \label{sec:triggering-policy}

Polarization compensation routines naturally define ``periods'' of protocol execution, which we define to be the time between the end of one compensation routine and the end of the next.
As an approximation to our objective of maximizing the long-run average entanglement distribution rate, we aim to maximize the average rate of each individual period.

For a given time $t$, let $\Bar{r}(t)$ be the average rate of the current period if we were to initiate compensation at time $t$.
Then we can show that the average rate of the period is maximized when the instantaneous rate equals the period's average rate, and therefore we should initiate a compensation whenever
\begin{equation}\label{eqn:rate-maximization}
    r(t) \leq \Bar{r}(t).
\end{equation}
See Appendix~\ref{sec:rate-max-derivation} for a derivation of this policy.

Since the rate is determined by our chosen pump power and the compensation duration is set to a constant, we can easily track the quantities $r(t)$ and $\Bar{r}(t)$.
We initiate a fidelity check whenever the condition in Equation~\ref{eqn:rate-maximization} is met to ensure that we do not operate in this suboptimal regime\footnote{Note that this is a second criterion to initiate a fidelity check, supplemental to the condition given in Section~\ref{sec:fid-check-policy}}.

Upon initiation, it is possible that $r(t)$ is simply suboptimal due
to our conservative choice of pump power.  Therefore, we set
$\Ftrigger$ to be the value of $\Ftrans$ that, if measured, would
result in a rate that violates the condition in
Equation~\ref{eqn:rate-maximization}. This is computed directly from
our pump power policy.

\subsection{Putting It All Together} \label{sec:all-together}
 
We explain how the components of our link control protocol work together through a representative run over a 29-second period between two polarization compensations, shown in Figure~\ref{fig:example-run}.
The following descriptions correspond to the circled numbers in Figure~\ref{fig:example-run}:

\begin{figure*}
    \centering
    \includegraphics[width=\textwidth]{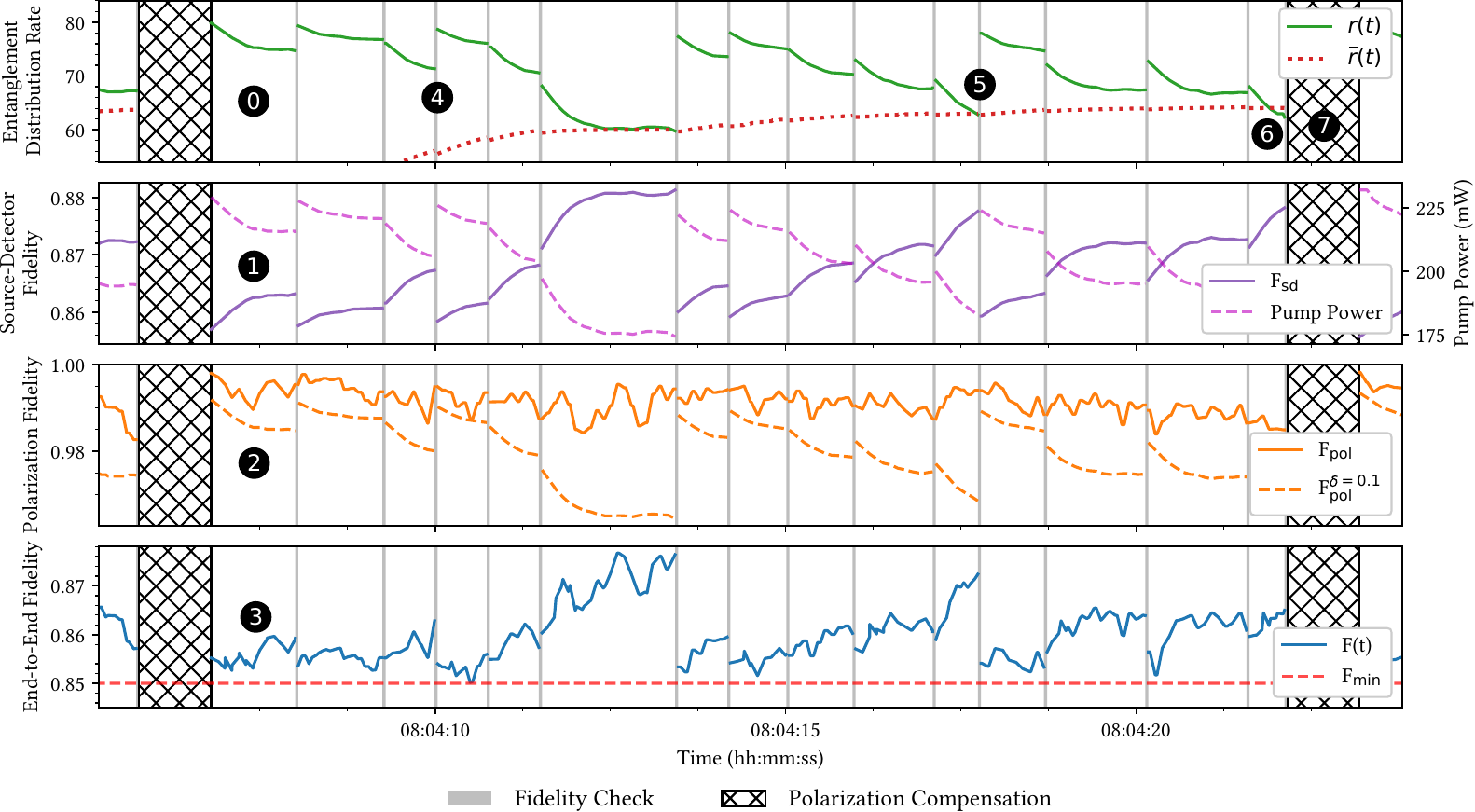}
    \caption{Example execution of the adaptive protocol across one period, showing all major characteristics. From top to bottom: end-to-end rate, pump power and source-detector fidelity, actual and predicted polarization fidelity, and end-to-end fidelity. See Section~\ref{sec:all-together} for a detailed explanation.}
    \label{fig:example-run}
\end{figure*}

\begin{enumerate}[leftmargin=*]
    \setcounter{enumi}{-1}
    \item The solid green curve in the topmost subplot shows the end-to-end entanglement distribution rate $r(t)$ at each time $t$. Notice that $r(t)$ decreases in the time between each fidelity check (thin vertical lines), and that $r(t)$ trends downward throughout the period overall.
    \item The second subplot shows pump power and $\Fsource$ over time. Pump power is chosen to satisfy the condition in Equation~\ref{eqn:conservative-pump-power}. Notice that the trends in $\Fsource$ mirror that of the pump power, while trends in $r(t)$ mimic the pump power. This stems directly from the experimental relationships between these quantities shown in Figure~\ref{fig:sd-characterization}.
    \item The third subplot shows the trends in actual and predicted polarization fidelity. After each fidelity check, our estimate $\Ftrans^{\delta=0.1}$ changes based on the observed drift since the previous check. $\Ftrans^{\delta=0.1}<\Ftrans$ throughout this example; in general, we expect this to be true at least $90\%$ of the time based on our choice of $\delta=0.10$.
    \item The bottom-most subplot shows the end-to-end fidelity as a result of the parameters from the plots above. At all times, the algorithm attempts to maintain an end-to-end fidelity above the minimum requirement $\Fmin$.
    \item Most fidelity checks occur due to the condition in  Equation~\ref{eqn:fid-check-condition}. In this instance, the protocol determines that the expected rate gain from performing a fidelity check outweighs the associated loss.
    \item To ensure $\Bar{r}$ is monotonically increasing, the protocol initiates a fidelity check whenever the instantaneous rate drops below the running average for the period (Equation~\ref{eqn:rate-maximization}). In this particular instance, the polarization prediction $\Ftrans^\delta$ was overly pessimistic, so $r(t)$ was unnecessarily low and a full compensation was not triggered.
    \item Once again $r(t)$ dips below $\Bar{r}(t)$, but this time the rate cannot be recovered above $r(t)$ and a polarization compensation is initiated. Notice that this event coincides with a substantial drop in $\Ftrans$.
    \item Each polarization compensation routine lasts a fixed duration of 1 second, as described in Section~\ref{sec:timeout-policy}.
\end{enumerate}

\section{Results\label{results}}

Here we present evaluations of our link control protocol, starting
with a discussion of how the results were derived.  Next, we compare
our protocol to currently adopted practices, and present results under
varying fidelity requirements as determined by higher level
applications.  Finally, we examine the sources of advantage in our protocol through a pair of intermediate baselines and we benchmark against a theoretical upper bound
which assumes no polarization drift.

\subsection{Methodology}

All results are derived from a custom trace-driven simulator. Source
code for this simulator and the underlying trace data are publicly available~\cite{source-code}.

The underlying trace data were derived over the 64
km partially-aerial LTS-NIST fiber link and span 68 hours (Figure~\ref{fig:polarization}). 
All evaluations were performed over the first 39 hours of this dataset, with the remainder reserved for drift model training.
Each data point
contains the Stokes parameters measured every 100 milliseconds,
enabling fine-grained reconstruction of polarization drift.  In our
simulations, we use the drift data to compute the polarization
fidelity $\Ftrans$ over time, which is then used to evaluate the
decisions made by the control algorithms.

Now consider a ``static'' protocol that issues an APC probe with some fixed
input parameters ($\Ftrigger,\Ftarget,\timeout$) every 5 seconds and
maintains constant pump power.  After each probe, $\Ftrans \geq
\Ftrigger$ by construction; polarization fidelity then degrades over
the next five seconds until the next probe. 
The end-to-end fidelity is given by the product of the fixed
source–receiver fidelity $\Fsource$ and $\Ftrans(t)$, and the
entanglement rate is derived directly from the pump power.  While
evaluating static protocols, we optimize over fidelity check frequency
and $\Fsource$ (a linear function of pump power, see
Figure~\ref{fig:sd-characterization}).  The values of the APC
parameters are fixed at $\Ftrigger=0.98, \Ftarget=0.99, \timeout=55s$,
consistent with a recent demonstration using the same
device~\cite{yicheng-entanglement-distribution}.

We apply the same method to evaluate our adaptive protocol, accounting
for dynamically adjusted parameters.  Pump power is updated every
100ms.  
APC probes are scheduled as needed based on the protocol state
(Sections~\ref{sec:fid-check-policy}-\ref{sec:triggering-policy}).

\subsection{Static Application Demands} \label{sec:results-static-demands}

\begin{figure*}[b]

    \vspace{.2cm}

    \begin{subfigure}{\linewidth}
        \includegraphics[width=\linewidth]{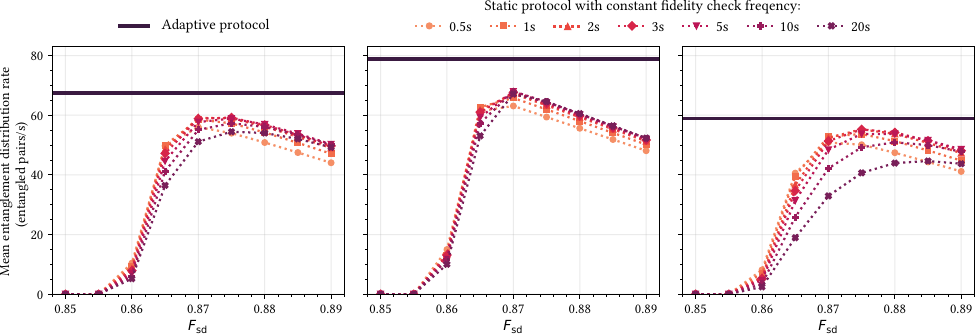}
    \end{subfigure}

    \hspace{.05\linewidth}
    \begin{subfigure}{.3\linewidth}
        \caption{Overall (24hr period)}
        \label{fig:results-overall}
    \end{subfigure}
    \hspace{.015\linewidth}
    \begin{subfigure}{.3\linewidth}
        \caption{Nighttime (10pm-6am)}
        \label{fig:results-night}
    \end{subfigure}
    \hfill
    \begin{subfigure}{.27\linewidth}
        \caption{Daytime (10am-6pm)}
        \label{fig:results-day}
    \end{subfigure}
    \hspace{.01\linewidth}

    \caption{Results comparing our adaptive link control protocol with
      static protocols (with different pump powers/$\Fsource$ and
      fidelity check frequencies) under static application demands ($\Fmin=0.85$).}
    \label{fig:daynight}
\end{figure*}

Figure~\ref{fig:daynight} compares our adaptive protocol to different
static protocols over different time intervals: 24 hours
(Figure~\ref{fig:results-overall}), 8 hours of night time
(Figure~\ref{fig:results-night}), and 8 hours of day time
(Figure~\ref{fig:results-day}).  Here, we set the application fidelity
demand ($\Fmin$) as a constant (0.85).  For the static protocols, we
vary both $\Fsource$ and fidelity check intervals. 

The result shows that the adaptive protocol performs well across all
time intervals, whereas static protocol parameters require careful selection for reasonable performance.  Over a 24 hour period, the
adaptive protocol outperforms the best static protocol by 14\%, and
this increases to 16\% at night.  The gap is smaller during the day (6\%), but it is
important to understand that the ``best'' static protocol for one day
could be very different for another.
Fidelity remains above $\Fmin$ 92.2\% of the time during the day, 99.4\% during the night, and 95.0\% across 24 hours.

Static protocol performance shows an expected trend: if the
source-detector fidelity $\Fsource$ is set too close to $\Fmin$, then
these protocols are not able to provide high entanglement throughput as nearly all pairs are below threshold.
Conversely, setting a more conservative $\Fsource$ does not allow the
static protocols to maximize entanglement rate.  Even with a well-chosen $\Fsource$, the best fidelity check interval varies over time (e.g., \mytilde1s during the day versus \mytilde10s at night).  Further, these parameter
values are sensitive to application demands ($\Fmin$) and the
polarization drift rate, both of which may vary unpredictably.
No parameter choices are required
for the adaptive algorithm.

\subsection{Varying Application Demands}

\begin{figure}[t]
    \centering
    \begin{minipage}{0.58\textwidth}
        \centering
        \includegraphics[width=\textwidth]{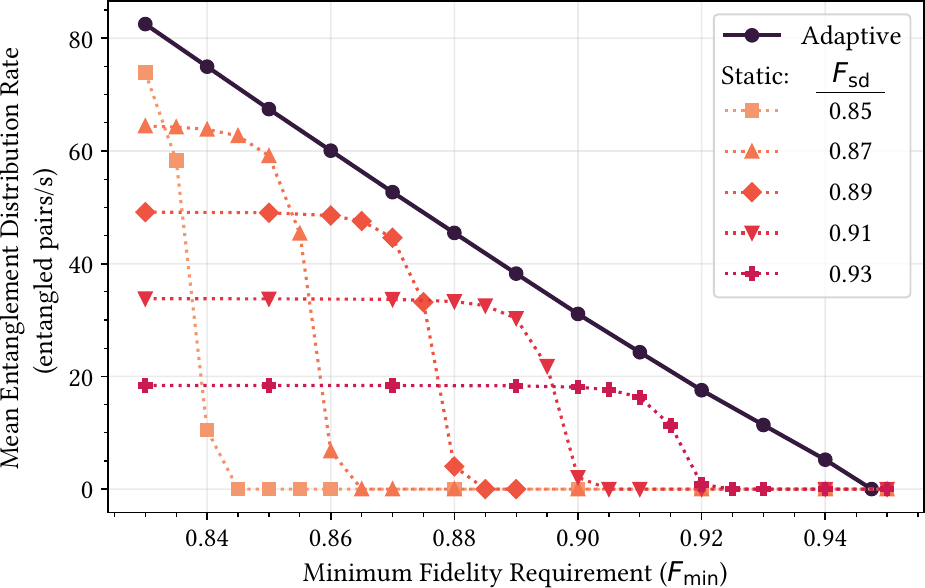}
        \caption{Comparison of adaptive protocol against static protocols with changing fidelity targets $\Fmin$.}
        \label{fig:fmins}
    \end{minipage}
    \hfill
    \begin{minipage}{0.38\textwidth}
        \centering
        \includegraphics[width=\textwidth]{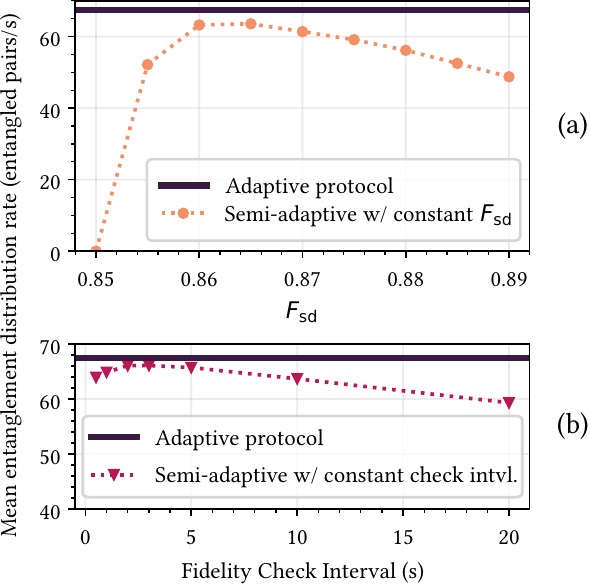}
        \caption{Intermediate baseline testing against partially-adaptive protocols.}
        \label{fig:intermediate-baselines}
    \end{minipage}
\end{figure}

The previous result fixed the application requirement of fidelity
($\Fmin$) to 0.85.  However, we expect a real-world deployment of
quantum networks will have to support multiple applications, each of
which may have different requirements.  One of the major
advantages of our adaptive protocol is that it can adapt to changing
$\Fmin$ ``on-the-fly''. 

Figure~\ref{fig:fmins} shows entanglement distribution rates achieved
($y$-axis) by the adaptive protocol and different static protocols
with changing $\Fmin$ ($x$-axis).  The results (obtained over a
24-hour run) again show that the adaptive protocol is able to
outperform each static protocol, but more importantly, each static
protocol optimizes for a single $\Fmin$.

The result also shows the limits of the adaptive protocol, as
entanglement rate steadily dips (and reaches 0 for
$\Fmin\approx0.9475$).  This is expected, as the higher values of
$\Fmin$ cannot be sustained by the adaptive (or indeed, any protocol),
as even the most conservative source setting (25mW pump power)
provides $\Fsource\approx0.9475$ (see
Figure~\ref{fig:sd-characterization}).

The previous result shows that the adaptive protocol is able to adapt
to different $\Fmin$ requirements (over 24 hours). However, the
protocol can also adapt to instantaneously changing $\Fmin$, as
$\Fmin$ is simply used as an input to
Equation~\ref{eqn:Fsd-calculation}.  We show the results from an
artificial scenario where the required $\Fmin$ changes continuously
in the shape of a sine curve.
While we do not expect application demands to necessarily change with
every entanglement requested, this experiment provides an ``extreme''
scenario to understand and visualize the dynamics of the adaptive
protocol.  It's important to note that the adaptive protocol is not
only adapting to the changing $\Fmin$, but also to the unknown link
drift simultaneously.  Finally, it is perhaps not a stretch to
emphasize that no static protocol would perform well in such a
scenario.

\begin{figure*}
    \centering
    \includegraphics[width=\textwidth]{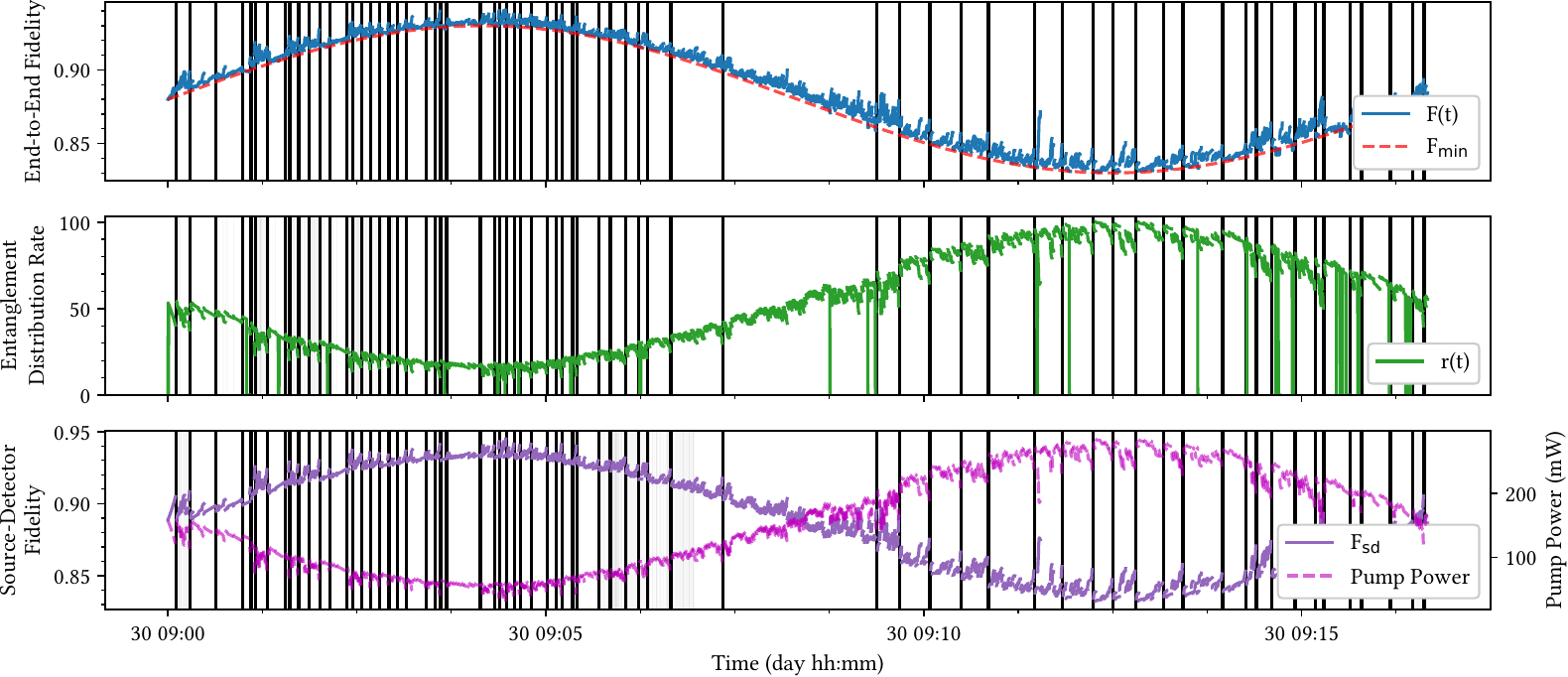}
    \caption{Protocol adaptivity under continuously-changing link demands following a sine curve (dashed red line): (top) end-to-end fidelity, (middle) entanglement distribution rate, (bottom) pump power and $\Fsource$. Vertical black lines indicate polarization compensations.}
    \label{sine}
\end{figure*}

Figure~\ref{sine} shows the performance of the adaptive protocol with
continuously changing $\Fmin$ (red curve in Figure~\ref{sine}a).
The figure shows that the adaptive protocol is able to maintain
requisite fidelity (blue line in Figure~\ref{sine}a) with changing
$\Fmin$.  Figures~\ref{sine}b and~\ref{sine}c show the entanglement rate
achieved and pump powers set by the protocol over time.  As expected,
the achievable entanglement rate mirrors $\Fmin$, as does the pump
power set.  It is interesting to note that the protocol is
compensating for {\em two} changing variables at each step here: polarization
drift over the link, and $\Fmin$.  We also see that the protocol
invokes the APC probes more frequently as we approach the maximum
feasible $\Fmin$, and less frequently as $\Fmin$ recedes.  This
behavior can be explained as follows: as $\Fmin$ approaches the maximum attainable $\Fmin$ with this
experimental setup, the protocol is frequently forced to set pump power to the
minimum available, triggering an
APC probe.  On the other hand, as the required $\Fmin$ decreases, the
instantaneous rate and average rate curves tend to not converge (See
Equation~\ref{eqn:rate-maximization} in
Section~\ref{sec:triggering-policy}), allowing for longer runs without
re-calibration.

\subsection{Sources of Performance Gain}

To better understand the nature of the performance gains achieved by the adaptive protocol, we compare against two semi-adaptive protocols.
These intermediate baselines are designed to decouple the relative contributions of adaptive APC triggering and adaptive pump power modulation.
Each semi-adaptive protocol uses the same parameter values as the fully adaptive version (e.g., $\timeout=1.0s, \Ftarget=1.0$) unless otherwise noted here.

First, we compare our full protocol against a semi-adaptive protocol with constant pump power (and therefore constant $\Fsource$) but adaptive APC scheduling. The APC is triggered each time the predicted polarization drift causes $F(t)<\Fmin$.
Figure~\ref{fig:intermediate-baselines}a shows the performance of this algorithm for various static values of $\Fsource$ over a 24 hour period, with mean entanglement rate about 6\% that of the fully adaptive protocol for $\Fsource=0.865$.
Notably, this adaptive-APC protocol outperforms all static protocols (see Figure~\ref{fig:results-overall}).
In particular, we observe significantly better performance for $\Fsource<0.87$, as the adaptive APC scheduling allows for the higher $\Ftrans$ required as $\Fsource$ approaches $\Fmin$.

Next we test the contribution of pump power adaptation by comparing against a protocol with adaptive pump power modulation and APC calls at constant intervals.
Figure~\ref{fig:intermediate-baselines}b shows that this protocol performs within 2.2\% of the fully adaptive scheme across 24 hours for fidelity check intervals around 2-3 seconds.
This suggests that the greatest source of performance gain in our protocol stems from increased link utilization due to dynamic pump power modulation.

\subsection{Evaluation Against Theoretical Upper Bound}

The adaptive protocol presented here provides no guarantees of
optimality, and indeed the process of defining an optimal protocol is
unclear.  Even if we consider an ``optimal'' that has perfect future
knowledge of drift, it is unclear how often to initiate APC probing,
and with what parameters.  If we further assume polarization
compensations take zero time, then the best policy would be to
continuously compensate.

In this ``optimal'' scenario, the system is equivalent to a
link with no drift, and pump power should be chosen
such that $\Fsource=\Fmin$. The throughput in the zero-drift scenario
therefore constitutes an (unachievable) upper bound on the performance
of any protocol.  We compare the performance of the adaptive protocol
to this theoretical upper bound.

For the experiments in Figure~\ref{fig:daynight}, this bound is a rate
of 85.6 entanglements/second. Over a 24-hour period, the adaptive
protocol performs within 21\% of the upper-bound (with \mytilde9\% of that
time being devoted to probes and compensation).  The comparison at
night is more interesting, as drift rate is lower and more closely
resembles the upper bound scenario.  At night, the gap is $<8\%$,  of which 3.4\% is
due to probes and compensation.  
These results indicate
that the scope for improvement for more sophisticated prediction and
compensation is relatively modest (<5\%).  
Moreover, the adaptive protocol is relatively simple and lightweight, and can easily be integrated into existing
deployments without major investment in computing hardware.

\section{Discussion} \label{sec:discussion}

\subsection{Deployment Considerations} \label{sec:discussion-deployment}

A practical advantage of the proposed protocol is that it requires only software access to control interfaces already common in deployed wide-area quantum links.
In the system studied here, the controller adjusts the source pump power via an electronically-tunable attenuator and adjusts the APC probe parameters through an API provided by the manufacturer.
Thus, deployment requires no additional quantum hardware.
It requires only a classical control process with access to existing control knobs

The online control logic is lightweight.
Runtime decisions consist of table lookups and $O(1)$ time parameter calculations.
The APC simulation used in our evaluation (Algorithm~\ref{alg:apc-sim-alg}) is not part of deployment, since compensation is executed directly by the APC device.
Similarly, the controller requires only low-bandwidth communication for distributing parameters and fidelity check results.
The relevant timescale in our simulations is 100ms; we expect that the lightweight computation, control, and communication requirements will fit well within this window.

The main deployment cost is characterization.
The entanglement source must have a well-characterized rate-fidelity curve and there must be sufficient polarization drift statistics to initialize the prediction model.
These measurements are standard characterization procedures that can be performed during commissioning and refined during normal operation.
This makes the protocol suitable for near-term deployment as a software control layer over existing wide-area quantum link infrastructure.

\subsection{Extension to General Quantum Links} \label{sec:discussion-generality}

While the control protocol presented here was developed around
specific quantum hardware, we believe that the fundamental concepts apply
broadly.  The problem classes \opt{0} (optimization of controllable
parameters) and \opt{1} (compensation of uncontrollable parameters)
are universal to all quantum links
and therefore \opt{2} (the joint optimization of \opt{0} and \opt{1}
over the same link) represents a general and practical class of
problems for which the methods presented here provide a blueprint.

\opt{0} formalizes the idea of managing tunable hardware parameters to
improve end-to-end system performance, rather than simply maximizing
the performance of individual components.  Quantum hardware often
exposes many parameters that can be tuned.
Many intrinsic hardware tradeoffs beyond pump power can be exploited for optimization; we
discussed several examples in Sections~\ref{sec:sources}
and~\ref{sec:receivers}. 
As a further example, consider the following
binary tradeoff: two successful measurement outcomes
occur with equal probability, but only one corresponds to
higher-fidelity
entanglement~\cite{lukin-telecom-memory-entanglement}.  This forces a
choice between discarding low-fidelity events to maximize fidelity or
retaining all events to maximize rate.

\opt{1} concerns compensation of uncontrollable noise parameters,
which is again universal.  Polarization stabilization is required for
all photonic quantum links regardless of encoding, and notably is
relevant for optical entanglement swapping, where increasing
polarization mismatch reduces success probability.  The magnitude and
timescale of polarization drift depend on link characteristics (e.g.,
length, aerial versus buried fiber), but every compensation strategy
induces tradeoffs.  Coexistence schemes that multiplex reference light
with quantum signals avoid system downtime but yield imperfect
compensation due to wavelength-dependent drift.  Conversely, using
same-wavelength reference signals enables high-precision compensation
but necessitates downtime as the reference signals overwhelm the
quantum signals.

Taken together, these considerations make \opt{2} a general problem
that applies to all wide-area quantum settings.  While the specific
control and response variables will vary by link, a common general
structure emerges: controllable parameters induce tunable tradeoffs
between entanglement rate and fidelity, while uncontrollable drift
introduces noise that must be mitigated at the cost of rate and/or
fidelity.  The core control-theoretic principle is to adjust
controllable parameters to offset the effects of drift while
compensating for uncontrollable parameters in a way that balances the
cost and benefit of compensation.  We expect these concepts to extend
naturally to a wide range of quantum links beyond the specific system
studied here.

\subsection{Extension to Higher Layers of the Quantum Networking Stack} \label{sec:discussion-opt3}

A natural extension of the above framework is the joint
optimization of two or more links working in tandem to distribute
entanglement over multiple hops. Each link has its own sets of
controllable parameters (\opt{0}) and time-varying noise parameters
(\opt{1}); we refer to the resulting multi-link problem as \opt{3}. 
Unlike earlier cases, \opt{3} must account for network topology while jointly configuring parameters and compensation strategies on a per-link basis.

Even for two links, \opt{3} is a broad and largely open problem when realistic hardware constraints are considered. 
Although out of scope for this paper, we illustrate the key considerations with a two-hop example. 
Consider three nodes $A$, $B$, and $C$,
connected by links $A$-$B$ (Link~0) and $B$-$C$ (Link~1),
with entanglement swapping at $B$~\cite{swap}. 
Suppose Link~0 is stable (e.g., short underground fiber), while Link~1 exhibits significant drift.  
An application specifies a minimum end-to-end entanglement quality requirement, which constrains the allowable operating points of each link.

Under an \opt{2} protocol, each link exposes a set of achievable rate–fidelity operating points.
Because Link~1 is more unstable, its feasible region is more restricted. 
An \opt{3} controller must then choose how Link~0 operates relative to Link~1: 
matching its rate, 
increasing fidelity to improve the end-to-end fidelity,
increasing its rate to enable entanglement distillation~\cite{distillation}, 
or adopting another strategy.
The optimal choice depends on hardware capabilities and how entanglement is ultimately consumed by higher layers.
This scenario demonstrates that \opt{2}(and our protocol) provides a useful abstraction for link behavior.
By exposing a feasible set of rate–fidelity operating points, \opt{2} grants higher-layer \opt{3} protocols flexibility in optimizing end-to-end performance.

Other practical challenges to scaling beyond a single link include compounding errors and timing constraints.
Fidelity decreases exponentially with hop count when swapping together many single-hop Bell pairs, which must be accounted for under \opt{3}.
Additionally, extending entanglement across multiple hops may impose timing constraints not considered under \opt{2}. For example, our protocol gives no guarantees on the timing of APC calls, so coordinating downtime across links may allow for more efficient multi-hop entanglement distribution.

Scaling is also likely to require entanglement distillation, where multiple entangled pairs are buffered and then consumed to create a smaller number of higher-fidelity pairs. As a link layer protocol, our protocol does not incorporate distillation, but should help regularize the outcome of distillation schemes by distributing entanglement at reliable fidelities.
In setting where the link generates multiple pairs intended for distillation, it may be desirable to alter the controller to optimize for a different metric that more closely connects to the ultimate distilled Bell pair fidelity, e.g., mean fidelity around a target value instead of a hard cutoff $\Fmin$.

\subsection{Limitations}

The design of the quantum link control protocol presented in this work involves several limiting characteristics which we hope to improve in future iterations.
First, the proposed control rules and evaluation are heuristic in nature and do not provide any formal optimality guarantees.
Second, hardware availability limited our evaluation to a single link, and some aspects of the protocol will need to change if other links are run with different technologies.
Third, we did not have access to a precise APC model, and real execution is likely to differ from our model (see Appendix~\ref{sec:appendix-apc-hist}).
Finally, our evaluation relies on trace-driven simulations without a closed-loop demonstration on a deployed system; significant depolarizing noise or intensity noise in the channel could change the rate-fidelity relationship and limit the efficacy of source tunability in an end-to-end experiment.

\section{Conclusions\label{conc}}

In this work, we presented a control protocol for quantum links that explicitly accounts for physical-layer impediments while exploiting controllable rate-fidelity tradeoffs in modern quantum networking hardware. 
Improvements in the performance of present-day quantum networks are typically driven by advances in hardware, yet our results demonstrate that carefully designed software control can yield substantial gains without increasing hardware requirements. 
Importantly, the proposed protocol has a clear path to deployment within existing experimental infrastructures. 
More broadly, the optimization framework introduced here extends beyond the specific link studied and highlights the potential for software-based control to play a significant role in improving the performance of near-term quantum networks.

\begin{acks}
We thank the anonymous reviewers and shepherd for their feedback which improved the clarity of the manuscript and strengthened the evaluation of our protocol.
We thank Mael Flament, Gabriel Bello Portmann, Rourke Sekelsky, and Aditya Verma of the Qunnect team for helpful discussions about the mechanics of the APC and for characterization data of the Qu-SRC.
We also thank Ivan Burenkov, Alan Migdall, and Thomas Gerrits for helpful conversations and comments during the preparation of the manuscript, and we thank Geng Liu for his work on the adaptive APC timeout algorithm.
This work was supported by a National Science Foundation Graduate Research Fellowship (Grant No. DGE 2236417).

\end{acks}

\bibliographystyle{ACM-Reference-Format}
\bibliography{reference}

\appendix

\section{Details of Entanglement Sources} \label{sec:source-detailed}

\begin{figure*}[b]
    \centering
    \includegraphics[width=.9\linewidth]{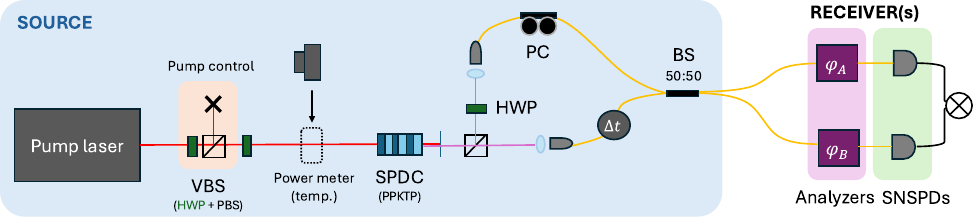}
    \caption{A more detailed diagram of the entanglement source sketched in Figure~\ref{fig:source-diagram-simplified}.}
    \label{fig:source-diagram-full}
\end{figure*}

This section provides additional detail on the SPDC setup beyond the simplified diagram introduced in Figure~\ref{fig:source-diagram-simplified}. As a general overview: the SPDC process occurs within some nonlinear media, whereby high-energy pump photons are converted into lower-energy, entangled photon pairs. Additional components are then used to harness entanglement in the desired degree of freedom. Each of these aspects is detailed below for our source.

The full experimental setup is shown in Figure~\ref{fig:source-diagram-full}. A mode-locked pump laser generates pulses at a repetition rate of 76.142 MHz.
Pump power control is executed via a variable beam splitter (VBS) consisting of an adjustable half wave plate (HWP) followed by a polarization beam splitter (PBS).
Adjusting the angle of the HWP changes the amount of pump light that used for SPDC.

The pump pulses then enter a periodically poled potassium titanyl phosphate (PPKTP) non-linear crystal which serves as a medium for the SPDC process. The crystal is engineered to convert pump photons with central wavelength 775 nm into photon pairs centered at 1550 nm.
With some probability ($\mytilde 10^{-4}$ per pulse), a pair of photons with orthogonal polarization states are generated, separated from the pump with spectral filters, and then split at a PBS. One of the photons passes through a HWP to (optionally) put both photons into the same polarization state.

The photons are coupled from free space into single-mode fiber and interfered at a 50:50 fiber-based beam splitter (BS)~\cite{shihalley_1988}. Interference is maximized through tuning of a variable delay ($\Delta t)$ to ensure photons arrive simultaneously, and a manual polarization controller (PC) to maintain photon polarizations in fiber.
Finally, the entangled photons propagate to the receiver. The analyzers set polarization bases $\left\{\varphi_A,\varphi_B\right\}$, and detection events at each SNSPD are timestamped with a time tagger (not shown) to count coincident detections. Through multiple measurements in different bases~\cite{tomography_2005}, the receiver fully reconstructs the state of the polarization-entangled photons.

\section{Details of Entanglement Source Modeling} \label{sec:sd-model-detailed}

\begin{figure}[b]
    \begin{subfigure}{.48\columnwidth}
        \includegraphics[width=\columnwidth]{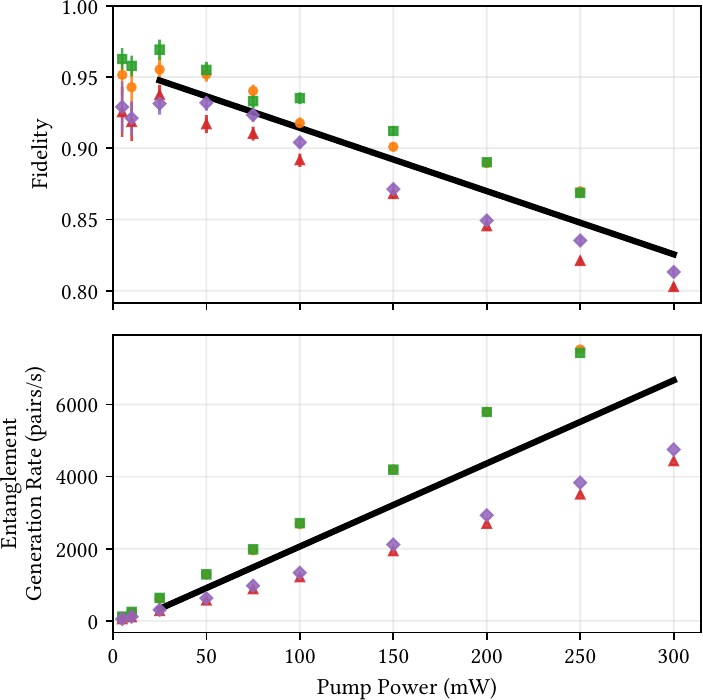}
        \caption{Full SPDC source characterization.\\\:
        \label{fig:sd-characterization-full}}
    \end{subfigure}
    \hfill
    \begin{subfigure}{.48\columnwidth}
        \includegraphics[width=\columnwidth]{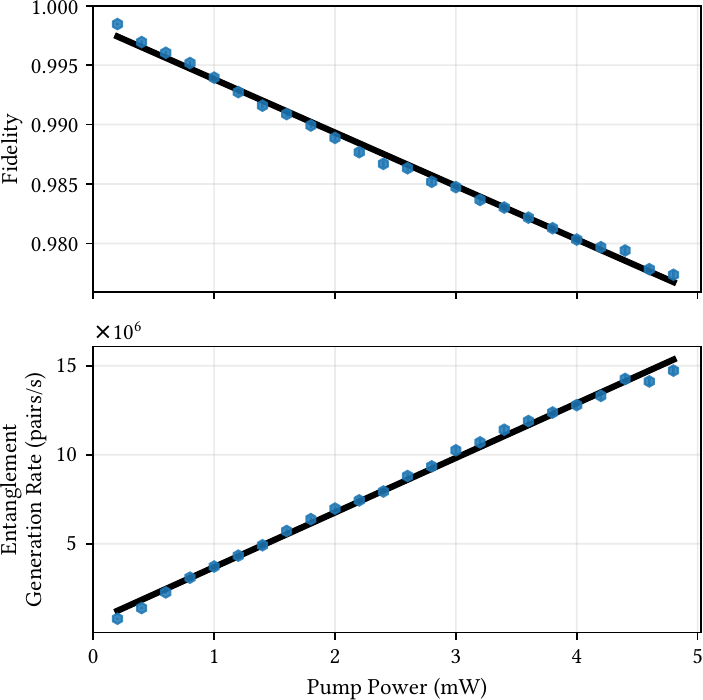}
        \caption{Full commercial four-wave mixing source characterization.
        \label{fig:sd-characterization-qusrc-full}}
    \end{subfigure}
    \caption{Extension of source characterizations in Figure~\ref{fig:tradeoffs} including all collected data.}
    \label{fig:tradeoffs-full}
\end{figure}

This Appendix includes more complete data used for SPDC and four-wave mixing source characterizations and development of the corresponding source models.

\subsection{SPDC Source}

Extending the main body plots from Figure~\ref{fig:sd-characterization}, additional information from the same data sets is shown in Figure~\ref{fig:sd-characterization-full}. This includes measurements at lower powers 5 mW and 10 mW, and raw count rates are shown without normalization. The four shown data sets correspond to measurement sequences taken on different days with slightly different conditions: green squares and orange circles on one day; purple diamonds and red triangles at a later date. Variations between trials are related to unstabilized polarization drift in the source-detector setup shown in Figure~\ref{fig:source-diagram-full}, which is elaborated upon below.

Entanglement generation rates are clearly consistent between data sets taken subsequent trials on the same day, but not on different days. This indicates different overall loss between source and receiver, largely attributed to the polarization-dependent SNSPD detection efficiency. The fibers between the analyzers and detectors are relatively stable over the course of one day, but not over longer time scales. The same is true (to a lesser extent) of the coupling efficiency from free space into fiber, which may have also reduced the overall efficiency in the later data sets. This combined effect simulates different link losses between source and receiver, so normalization is used to represent the underlying source dependence on pump power in Figure~\ref{fig:sd-characterization}.

State fidelity is also largely consistent between data sets, but with some increased variability between measurements taken on the same day. This is because entanglement is generated from quantum interference at a 50:50 beamsplitter, which is highly sensitive to the polarization of the incoming photons --- more sensitive than the SNSPD efficiency dependence. We characterize the quality of interference  with the Hong-Ou-Mandel (HOM) effect~\cite{hom1987}, observing about 96\% and 93\% visibility for the two different days. This value is limited by the ability to manually match polarization states in fiber via polarization controller (PC). This in turn limits the state fidelity at low powers, as seen at the top of Figure~\ref{fig:source-diagram-full}, and is also affected by drift, causing higher variability between trials.

Altogether, rates exhibit a clear linear trend over all measured pump powers, while fidelity shows no significant change until around 25 mW (for a given data set), above which multipair effects become significant and degrade fidelity. This defines an operable regime (25 mW to 300 mW) over which decreasing rate returns a consistent increase in state fidelity, offering a useful tradeoff that can be leveraged in our control protocol.

\subsection{Four-Wave Mixing Source}

In Figure~\ref{fig:sd-characterization-qusrc} we presented source characterization data for a Qunnect Qu-SRC~\cite{qu-src} which generates pairs of polarization-entangled photons via four-wave mixing (FWM) in warm atomic vapor.
We received this characterization data directly from the device manufacturer. The full (non-normalized) data is presented in Figure~\ref{fig:sd-characterization-qusrc-full}.

Note the given SPDC rates include all optical losses incurred by the apparatus in Fig.~\ref{fig:source-diagram-full}, while this commercial FWM source is highly optimized.
As a result, the commercial source emits photonic Bell states at a rate several orders of magnitude higher and an error rate one order of magnitude lower than our in-house SPDC-based source. 
The output photons also have much different wavelength characteristics: the SPDC source emits degenerate photons with the same center wavelength (1550 nm); the FWM source emits non-degenerate photon pairs with wavelengths 795 nm and 1324 nm, each with a much narrower spectral band.
These spectral properties will result in completely different polarization drift due to the dispersive properties of optical fiber. To translate our model to a new source, drift characterization similar to Fig.~\ref{fig:polarization} would be required for the relevant wavelengths, although we can expect similar dependence of drift rate on environmental conditions.

\section{Sensitivity of Polarization Drift Model Training} \label{sec:appendix-drift-model}

\begin{figure}[b]
    \centering        
    \includegraphics[width=.6\columnwidth]{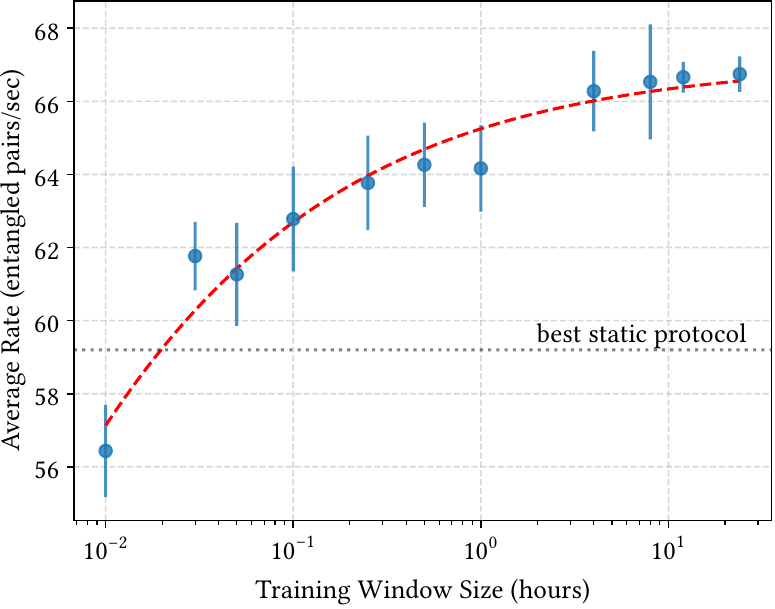}
    \caption{Effect of polarization drift model training duration.}
    \label{fig:training-window}
\end{figure}

The polarization drift model used to generate the results in this paper was trained on 24 hours of polarization drift data, with samples taken every 100ms.
The specific range of data used for training was 6am October 1 to 6am October 2 (see Figure~\ref{fig:polarization}), chosen so it does not overlap with the data used for trace-driven evaluation.
Drift measurements in this dataset were taken at 100ms intervals.

We evaluated the performance of our protocol using models trained on varyingly-sized contiguous segments of the training set.
Figure~\ref{fig:training-window} shows mean performance over a 24 hour period as a function of the training period length.
Each point is the mean of four trials run with models trained on data of the specified window size, with a log-logistic fit. Error bars show standard error.
We find a slow decline in performance as training window size decreases below 24 hours.
Performance remains above that of the best static protocol (59.2, see Figure~\ref{fig:results-overall}) until the training window size is reduced to on the order of 1 minute.
We also see plateauing performance for training window sizes on the order of a few hours and above.
This suggests that our chosen 24 hour training period is more than sufficient for our protocol, and a window one or two orders of magnitude smaller may also suffice.

An important aspect of a well-trained drift model is its ability to make accurate predictions across a wide range of drift conditions, as the link control protocol relies on accurate responses to frequently changing polarization drift behavior.
The training phase should therefore incorporate data that encompasses the full range of drift magnitudes that will be experienced at runtime.
This helps explain the performance saturation as we approach a 24 hour training window, as our training data includes a wide range of conditions across day and night.
In particular, our training dataset encompasses the full range of relevant drift conditions, from near zero drift at points during the night to drift magnitudes beyond the correction ability of the APC at points during the day.

It is worth noting that the polarization drift data required to build the model can be gathered directly by the APC, so the training requires no additional hardware setup.
Furthermore, drift measurements during protocol execution can be used to further train the drift model, potentially allowing for a dynamic model that adapts to longer-term changes in drift behavior.

\section{Fit of APC Model to Experimental Data}
\label{sec:appendix-apc-hist}

\begin{figure}[b]
    \centering        
    \includegraphics[width=.6\columnwidth]{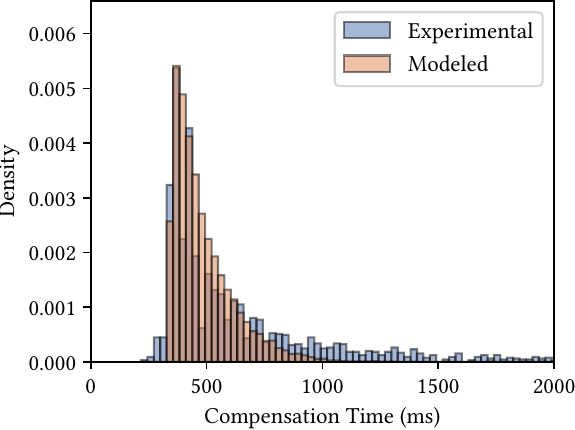}
    \caption{Histograms of APC compensation times generated by Algorithm~\ref{alg:apc-sim-alg} and experiment.}
    \label{fig:apc-times-hist}
\end{figure}

Our empirically-derived model of polarization compensation in the Qunnect APC, presented in Section~\ref{sec:apc-model}, assumes Riemannian gradient descent and a noise term that counteracts the improvements gained in each gradient descent step.
We fit our model to 70 hours of experimental data over a similar link and determined a gradient descent step size of $\eta=0.062$ and a step duration of $\Delta=27.7$ms.
Figure~\ref{fig:apc-times-hist} compares the histogram of compensation durations from the experimental APC dataset with a histogram derived from our simulations, which use our model on the data shown in Figure~\ref{fig:polarization}.

Visual comparison demonstrates reasonable empirical fit. 
However, there are some differences between the distributions, particularly at higher compensation times,
which are very likely due to discrepancies between the modeled APC procedure in Algorithm 1 and the actual algorithm run on the device. We made every effort to create a model that approximates the true performance of the APC, including explaining our modeling to the software team at Qunnect (the APC manufacturer). The Qunnect team verified that our model captures the right characteristics, but they could not disclose the details of their algorithm. It is likely that the true APC algorithm does not exhibit the behavior of perfect Riemannian gradient descent in the presence of drift. Although it would be possible to artificially tweak Algorithm~\ref{alg:apc-sim-alg} such that its performance matches that of this particular experimental dataset (blue bars in Figure~\ref{fig:apc-times-hist}), this approach would amount to an empirical overfitting of our model without more information of the proprietary routine.

\section{Derivation of End-to-End Entanglement Fidelity}\label{sec:appendix-fidelity}

For our dynamic control protocol described in Section~\ref{sec:protocol}, we model the end-to-end fidelity as a simple scalar product of two quantities: $F(t)\approx \Ftrans(t)\Fsource(t)$. Here, we derive a more complete expression for $F(t)$ and provide physical justification for this approximation.

The source-detector fidelity $F_{sd}$ is obtained empirically as a function of pump power, and represents the end-to-end fidelity in the absence of polarization drift. Physically, the imperfections introduced by the source and detectors can be modeled by a \textit{depolarizing} channel $\mathcal{D}$. The action of this channel on a quantum state with density operator $\hat{\rho}$ is shown below~\cite{Nielsen_Chuang_2010}:
\begin{equation}\label{eqn:depol}
    \mathcal{D}(\hat{\rho},\lambda)=(1-\lambda)\hat{\rho}+\lambda\frac{\hat{I}}{D}
\end{equation}
where $D$ is the Hilbert space dimension and $\hat{I}$ is the identity operator; for single qubits, $D=2$, and for entangled two-qubit Bell states, $D=4$. The second term in Eq.~\ref{eqn:depol} represents the addition of completely uncorrelated noise (a maximally mixed state) in place of the original state $\hat{\rho}$--- e.g., dark count noise from the detectors, or multi-pair emission from the source--- with probability $\lambda\in[0,1]$. Note that if depolarization is applied to both qubits of the Bell state, the result can still be modeled as a single depolarization channel with a combined effective parameter $\lambda$ as above~\cite{Shi_2025}.

Next, we find $F_{sd}$ by computing the fidelity of a pure input state $\hat{\rho}=|\psi\rangle\langle\psi|$ with the output of Eq.~\ref{eqn:depol}. For source-detector fidelity:
\begin{align}
    F_{sd}(\lambda)&=\langle\psi| \mathcal{D}(\hat{\rho},\lambda) |\psi\rangle \nonumber \\
     &=(1-\lambda)\langle\psi| \hat{\rho} |\psi\rangle + \frac{\lambda}{D} \langle\psi| \hat{I} |\psi\rangle \nonumber \\
     &=(1-\lambda)+\frac{\lambda}{D} \label{eqn:fidsd}
\end{align}
where each of the matrix elements $\langle\psi|\cdot|\psi\rangle$ evaluates to $1$.

The end-to-end fidelity $F(t)$, the main quantity of interest, combines both depolarization $\mathcal{D}$ and unitary rotation $\mathcal{R}$-- i.e., polarization drift by some angle $\theta$. Using the fact that depolarization is invariant under unitary rotation~\cite{Holevo2006Additivity}, the order of $\mathcal{D}$ and $\mathcal{R}$ here is arbitrary, and so we can repeat the same calculation as above, only using the rotated input state $\mathcal{R}(\hat{\rho},\theta)$:

\begin{align}
    F(\theta,\lambda)&=\langle\psi| \mathcal{D}\left[\mathcal{R}(\hat{\rho},\theta),\lambda\right]|\psi\rangle \nonumber \\
     &=(1-\lambda)\langle\psi| \mathcal{R}(\hat{\rho},\theta) |\psi\rangle + \frac{\lambda}{D} \langle\psi| \hat{I} |\psi\rangle \nonumber \\
     &=(1-\lambda)\Ftrans(\theta)+\frac{\lambda}{D} \label{eqn:fide2e}
\end{align}
where $\Ftrans(\theta)$ is given exactly by Eq.~\ref{eqn:angle2fid} via Ref.~\cite{jozsa_1993}. With some algebra, the final expressions in Eqs.~\ref{eqn:fidsd} and~\ref{eqn:fide2e} can be combined to relate all three fidelity quantities:

\begin{equation} \label{eqn:Ftrue}
    F=\Fsource\Ftrans + \frac{1}{D-1}(1-\Fsource)(1-\Ftrans)
\end{equation}
As the second term is strictly positive, this expression is bounded below by the product $F\approx\Fsource\Ftrans$, and this approximation holds when both fidelities are close to $1$. For $\Fsource,\Ftrans\geq0.80$, this results in less than a 1.7\% relative error (underestimation) of the ``true'' end-to-end Bell state fidelity.

\begin{figure}[t]
    \includegraphics[width=.9\linewidth]{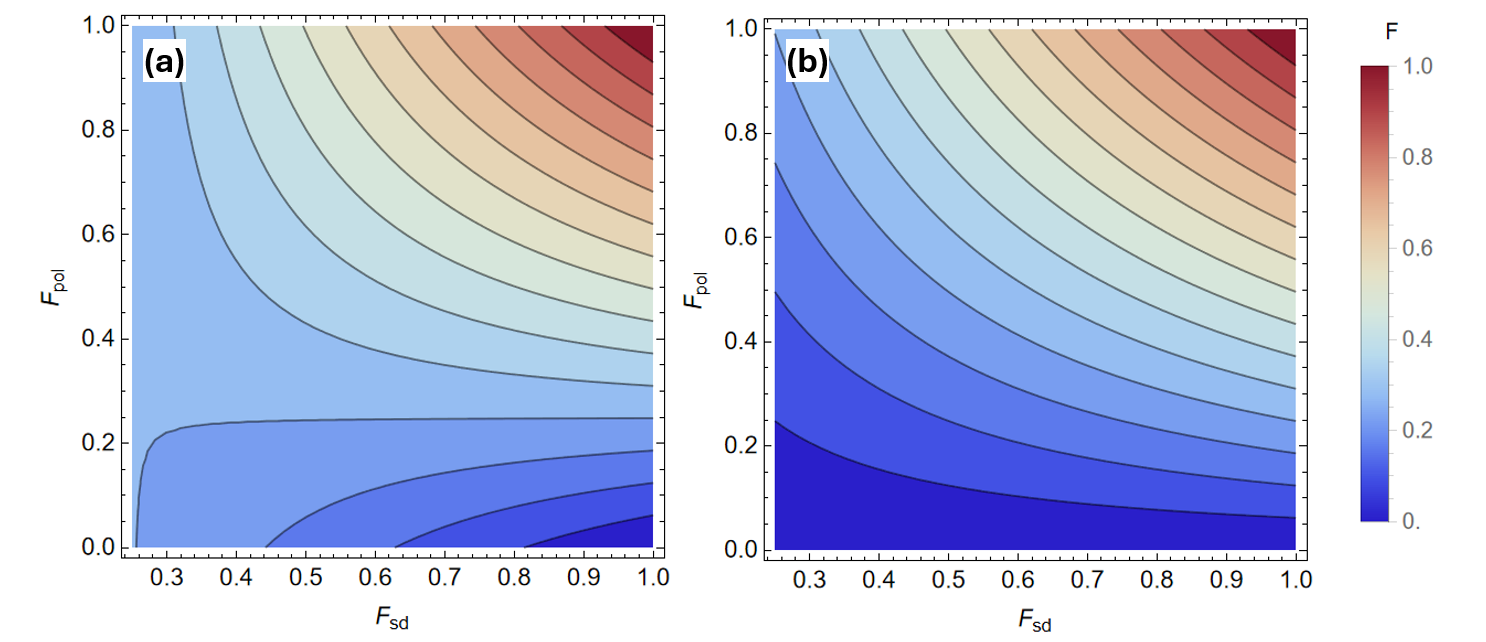}
    \caption{End-to-end fidelity $F$ for pure Bell state inputs plotted as a function of component fidelities $\Fsource$ and $\Ftrans$, using (a) the full expression (see Eq.~\ref{eqn:Ftrue}) and (b) the approximation $F\approx\Fsource\Ftrans$.
    \label{fig:fidmaps}}
\end{figure}

As seen in Fig.~\ref{fig:fidmaps}, the more detailed end-to-end fidelity model only meaningfully deviates from the simple approximation for very low values of $\Fsource$ and $\Ftrans$. This leads to a curious region ($\Ftrans<1/4$) in panel (a) where increasing $\Fsource$ actually \textit{decreases} the end-to-end fidelity, which appears to contradict our usual strategy. However, this is somewhat misleading for two reasons. First, this region lies well below any practical value of $\Fmin$, and so even in cases of extreme drift, compensation will always be applied to increase $\Ftrans$ rather than attempt to change $\Fsource$. Second, this non-monotonicity is a consequence of depolarized noise having nonzero fidelity with Bell states, despite having zero useful entanglement. This behavior is a limitation of the fidelity metric--- it is an imperfect proxy for the quality of entanglement and provides a slight ``false positive'' response to noise in this regime. In any case, our protocol avoids this by always applying compensation to increase $\Ftrans$ (and thus $F$) in low-fidelity scenarios, after which the typical behavior applies.

\section{Comparison with Classical Adaptive Control} \label{sec:appendix-classical-comparison}

An obvious point of comparison of our control protocol is how TCP chooses send window size in reaction to congestion. Classic TCP employs a Additive-Increase-Multiplicative-Decrease (AIMD) scheme to control source rate (modern versions use different variations), and many modern TCPs can also use in-network signals such as ECN. There is no ``competing'' traffic in our scenario, nor do we have a mechanism for signaling environmental changes. Thus, adaptive measures such as AIMD are not required or appropriate in our scenario. This is amplified by the fact that if the entanglement fidelity falls below the requisite minimum, then the utility of transmissions fall to zero due to our assumptions. Instead, we use periodic in-band measurements to estimate fidelity, and then rely on the pre-trained model to predict fidelity between measurements. The actual rate selection is a table lookup based on the source characterization done offline.

Another natural setting is to compare our scheme to existing schemes for rate control over wireless links. Classical wireless links choose different coding and/or modulations based on environmental conditions, e.g., they may reduce bit-depth (in QAM), or integrate over time (use redundant coding to increase symbol time) to decrease bit-error rate. Classic protocols may also use FEC or ARQ to decrease bit-error rate. Unfortunately, none of these schemes apply to quantum links, where the information is carried in individual photons. There is no notion of bit depth, nor can the modulation integrate over time/multiple photons. (Note that it is possible to employ entanglement distillation, which does integrate over multiple photons, but we see this as an option at a layer higher than the basic link layer). In a way, our protocol highlights the unique challenges associated with quantum links where classical approaches such as changing modulation or integration time do not apply, at least at the link layer. The approach we adopt is to try to predict the current link fidelity and then produce individual photons with sufficient quality that they may traverse the link at its current (predicted) state.

\section{Computing the Fidelity Check Condition}
\label{sec:appendix-rate-integration}

Here we discuss the computation of the quantity $\Lambda_1-\Lambda_2$ introduced in Section~\ref{sec:fid-check-policy}.
$\Lambda_1-\Lambda_2$ quantifies the throughput gain of performing a fidelity check:
$\Lambda_1$ is the expected throughput if we perform the check, and $\Lambda_2$ is the throughput if we don’t perform the check.

Let $e$ denote the time elapsed since the last fidelity check and $\Teff$ the expected time until the next fidelity check.
Let $r(\Ftrans)$ denote the rate achieved when the polarization fidelity is predicted to be $\Ftrans$. Let $t_\text{check}$ denote the duration of a fidelity check.

To ease the notation in this discussion, we define $\dot{r}(t):=r(t_0+t)$ where the most recent fidelity check occurred at time $t_0$ --- i.e., $\dot{r}(t)$ is the rate $t$ seconds after the last check.

The first important realization is that we can always compute the rate $\dot{r}(t)$ for $t>0$ up until the next fidelity check.
Intuitively this must be true: we learn no information about the system between fidelity checks, so we should be able to predict all protocol behavior until the next check.
Indeed, $\dot{r}(t)$ is determined by the pump power, which in turn is determined by our prediction of $\Ftrans$ at time $t$ (see Section~\ref{sec:power-policy}), so we can project $\dot{r}(t)$ by querying our polarization drift prediction model.
We employ this rate projection to estimate $\Teff$ by finding the time at which the condition in Equation~\ref{eqn:rate-maximization} is met and taking the minimum of this quantity and the time since the last check.

Given this, $\Lambda_2$ is easy to compute by noticing that 
\begin{equation}
    \Lambda_2 = \int_{e}^{e+\Teff}\dot{r}(t)dt
\end{equation}

Next, we estimate $\Lambda_1$.
We assume that the rate curve $\dot{r}_{\text{after-check}}(t)$ after a hypothetical fidelity check would follow the same trend as the current $\dot{r}(t)$ with an offset depending on the expected true polarization fidelity $\mathbb{E}[\Ftrans]$.
We compute $\mathbb{E}[\Ftrans]$ by querying our drift prediction model with $\delta=0.5$ ($50^{\text{th}}$ percentile drift).
We can now estimate $\Lambda_1$:
\begin{align}
\Lambda_1
&=\int_{0}^{\Teff-t_\text{check}}\dot{r}_{\text{after-check}}(t)dt \\
&\approx\int_{0}^{\Teff-t_\text{check}}\dot{r}(t)-\big[\dot{r}(0)-\dot{r}(\mathbb{E}[\Ftrans])\big]dt
\end{align}

This uses the same projection of $\dot{r}(t)$ that we used to compute $\Lambda_2$.
Note that we only need to perform the calculations here once per fidelity check.

\newcommand{\nent}{n_{\text{entanglements}}}

\begin{algorithm}[b]
\caption{Adaptive APC Timeout Selection}
\label{alg:adaptive-timeout}
\begin{algorithmic}[1]
\Require the current polarization fidelity $\Ftrans^0$; set of candidate timeout values $T_\text{options}$ (initially \{1.0\})
    \Ensure compensation duration $\timeout$

    \State $(T_{\text{best}}, r_{\text{best}}) \gets (0,0)$
    \Comment{Best timeout value so far and its rate}
    \State $dt \gets 0.1$ \Comment{100ms timesteps}
    \For{$\timeout \in T_\text{options}$}
        \State $t, r, \Bar{r}, \nent \gets 0$
        \State $\hat{F}_{\text{pol}}^0 = \Call{simulate\_apc}{\timeout, \Ftrans^0}$ \label{line:sim-apc}
        \Comment{Estimated $\Ftrans$ after running APC, via Algorithm \ref{alg:apc-sim-alg}}
        \While{$r \geq \Bar{r}$}
            \State $t \gets t + dt$
            \State $\Ftrans^t \gets \Call{drift\_predictor}{t, \hat{F}_0}$
            \Statex \Comment{Predict the polarization fidelity $t$ seconds after compensation ends} \label{line:drift-prediction}
            \State $\Fsource^t=\Fmin/\Ftrans^t$
            \Comment{Equation~\ref{eqn:Fsd-calculation}} \label{line:Fsd}
            \State $r \gets \texttt{rate}(\Fsource^t)$
            \Comment{Entanglement rate corresponding to $\Fsource$} \label{line:calc-rate}
            \State $\nent \gets \nent + r \cdot dt$
            \Comment{Update total throughput} \label{line:nent}
            \State $\Bar{r} \gets \nent / (t + \timeout)$
            \Comment{Compute average rate through time $t$} \label{line:rbar}
        \EndWhile
        \If{$r>r_\text{best}$}
        \State $(T_{\text{best}}, r_{\text{best}}) \gets (\timeout,r)$
        \EndIf
    \EndFor
    \State \Call{update\_timeout\_options}{$T_\text{options}, T_\text{best}$}
    \State \Return $T_{\text{best}}$

\end{algorithmic} 
\end{algorithm}

\begin{algorithm}[b]
\caption{Dynamically Set Timeout Options}
\label{alg:adaptive-timeout-options}
\begin{algorithmic}[1]
\Require $T_\text{options}, t_\text{picked}$
\Ensure compensation duration $\timeout$

\Function{update\_timeout\_options}{$T_\text{options},t_\text{picked}$}
\State $n_\text{timeouts} \gets 32$
\Comment{Maximum size of $T_\text{options}$}

\If{$t_\text{picked} \in \{\min(T_\text{options}), \max(T_\text{options})\}$} \Comment{If an extreme value is selected\dots} \label{line:if-min-max}

    \While{$|T_\text{options}| > n_\text{timeouts}-2$}
        \State $T_\text{options} \gets T_\text{options} \setminus \{\Call{least\_recently\_used}{T_\text{options}}\}$
        \Comment{Ensure $|T_\text{options}|\leq n_\text{timeouts}$} \label{line:ntimeouts}
    \EndWhile
    
    \If{$t_\text{picked} = \min(T_\text{options})$}
        \State $T_\text{options} \gets T_\text{options} \cup \left\{\frac{t_\text{picked}}{2}\right\}$ \Comment{Add a smaller value if it's the minimum} \label{line:min}
    \EndIf

    \If{$t_\text{picked} = \max(T_\text{options})$}
        \State $T_\text{options} \gets T_\text{options} \cup \{2t_\text{picked}\}$ \Comment{Add a larger value if it's the maximum} \label{line:max}
    \EndIf

    \State $t_{-1} \gets \max \{x \in T_\text{options} : x < t_\text{picked}\}$
    \State $t_{+1} \gets \min \{x \in T_\text{options} : x > t_\text{picked}\}$
    \State $T_\text{options} \gets T_\text{options} \cup \{\frac{t_{-1}+t_{+1}}{2}\}$
    \Comment{Add an intermediate value} \label{line:avg-update}
\EndIf
\EndFunction

\end{algorithmic} 
\end{algorithm}

\section{Dynamic Compensation Duration} \label{sec:appendix-dynamic-compensation}

In Section~\ref{sec:timeout-policy}, we chose to set all compensation durations equal to a constant 1 second.
Here we consider a dynamic compensation time selection algorithm which shows that our 1s constant selection is not optimal.
However, we find that this algorithm contributed substantial runtime overhead for minimal performance benefit, and therefore elect to use the static policy presented in Section~\ref{sec:timeout-policy}.

The dynamic compensation selection algorithm consists of two main subroutines: 
(1) create a set of candidate values, out of which the timeout value $\timeout$ will be chosen, and (2) simulate the performance of each candidate timeout and select the one with the highest expected average rate over the upcoming period.

We will describe subroutine (2) first.
Suppose for now we have already decided on a set of candidate timeout values $T_\text{options}=\{T_1, T_2, \dots, T_n\}$.
Algorithm~\ref{alg:adaptive-timeout} simulates the performance of each of these values. 
For each candidate, the first step in this simulation is to estimate the polarization fidelity that we will achieve after compensation ($\hat{F}^0_\text{pol}$, line~\ref{line:sim-apc}).
The while loop then simulates the next period's execution following this hypothetical APC compensation, terminating when the rate $r$ drops below the average rate $\Bar{r}$ according to the condition in Equation~\ref{eqn:rate-maximization}.
For each time $t$ in the next period, the while loop simulates the following: compute the expected drift at time $t$ (line~\ref{line:drift-prediction}), calculate the rate that the protocol would achieve given that drift prediction (lines~\ref{line:Fsd} and~\ref{line:calc-rate}), then compute the predicted average rate up to time $t$ (lines~\ref{line:nent} and~\ref{line:rbar}).
The algorithm returns with the timeout value that gives the greatest expected return, and updates the timeout candidates based on this selection.

Next we describe subroutine (1) for updating the set of candidate timeout values, given in Algorithm~\ref{alg:adaptive-timeout-options}.
Suppose the most recent selected timeout value was $t_\text{picked}$.
We only update the possible timeout options when $t_\text{picked}$ was either the smallest or largest available (line~\ref{line:if-min-max}).
If it was the smallest, we add a candidate value of $t_\text{picked}/2$ (line~\ref{line:min}), and if it was the largest, we add a candidate value of $2t_\text{picked}$ (line~\ref{line:max}).
We also then add the average of the two elements immediately smaller and larger than $t_\text{picked}$ (line~\ref{line:avg-update}), which will increase the resolution of the timeout options for the next iteration.
In this way, we adaptively expand the range of timeout candidates and granularity within that range during execution as necessary.
To keep the computational cost reasonable, we limit the number of candidate timeout values to $n_\text{timeouts}=64$ (line~\ref{line:ntimeouts}).
Tests with $n_\text{timeouts}<32$ also showed comparable results.

\begin{figure*}[t]
    \centering
    \includegraphics[width=.6\linewidth]{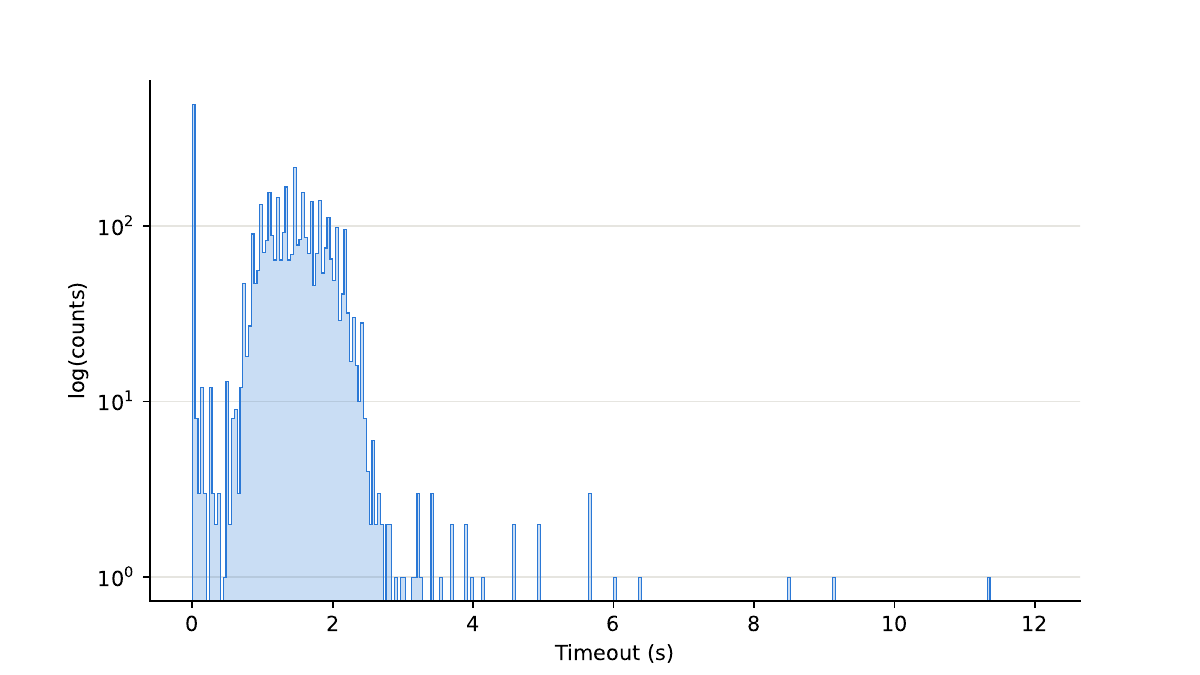}
    \caption{Distribution of APC timeout values dynamically selected by Algorithm~\ref{alg:adaptive-timeout} across a 24 hour simulation. The majority of selections fall into a window around 1-2 seconds, consistent with the optimal constant values in Figure~\ref{fig:dynamic-timeout}.}
    \label{fig:adaptive-timeout-hist}
\end{figure*}

\begin{figure*}[t]
    \vspace{.2cm}

    \begin{subfigure}{\linewidth}
        \includegraphics[width=\linewidth]{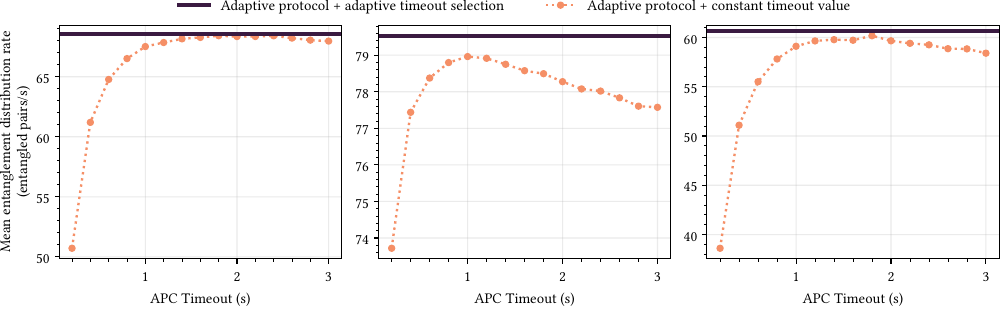}
    \end{subfigure}

    \hspace{.05\linewidth}
    \begin{subfigure}{.3\linewidth}
        \caption{Overall (24hr period)}
        \label{fig:dynamic-timeout-overall}
    \end{subfigure}
    \hspace{.015\linewidth}
    \begin{subfigure}{.3\linewidth}
        \caption{Nighttime (10pm-6am)}
        \label{fig:dynamic-timeout-night}
    \end{subfigure}
    \hfill
    \begin{subfigure}{.27\linewidth}
        \caption{Daytime (10am-6pm)}
        \label{fig:dynamic-timeout-day}
    \end{subfigure}
    \hspace{.01\linewidth}

    \caption{Performance of the adaptive link control protocol with the dynamic compensation duration algorithm in Appendix~\ref{sec:appendix-dynamic-compensation} compared to the adaptive control protocol with varying constant compensation durations. The simulation conditions are the same as in Section~\ref{sec:results-static-demands}.}
    \label{fig:dynamic-timeout}
\end{figure*}

The adaptivity of these procedures is seen in Figure~\ref{fig:adaptive-timeout-hist}, which shows a histogram of selected timeout values across a 24 hour run.
From this plot, we see that the algorithm mostly selects values in the range of approximately 1 to 2 seconds, as we would expect based on the results in Figure~\ref{fig:dynamic-timeout} (discussed next).
Sometimes the algorithm selects a timeout value of 0, typically when the polarization fidelity is already very high, in which case there is no effect.

We also evaluated this adaptive timeout selection procedure under the same conditions as in Section~\ref{sec:results-static-demands}.
The results are shown in Figure~\ref{fig:dynamic-timeout}. Each plot also shows the performance of the protocol under various constant timeout values.
Across day, night, and the 24 hour run, the adaptive timeout selection gives higher mean throughput than any constant timeout value.
The performance improvements seen (over the best constant values) are 0.2\% across 24 hours, 0.7\% during the night, and 0.8\% during the day.

After evaluating this adaptive APC runtime approach, we concluded that the simple constant scheme presented in Section~\ref{sec:timeout-policy} is preferable for the purposes of this evaluation.
While we see consistent benefits from dynamic timeout selection, the benefits are marginal and far outweighed by the high computational cost of Algorithm~\ref{alg:adaptive-timeout}.
In a real time quantum link controller, the cost of simulating many future execution paths before each APC call
appears impractical.
However, searching a more computationally efficient timeout selection algorithm may be an interesting topic for future research.

\section{Derivation of Optimal Compensation Timing} \label{sec:rate-max-derivation}

Here we derive Equation~\ref{eqn:rate-maximization}, which states the condition under which a protocol should initiate a polarization compensation routine. 
First, we approximate the objective of maximizing end-to-end rate by a greedy strategy of maximizing the rate of each period individually (from the end of one calibration routine to the end of the next).

Suppose each compensation routine takes time $c$, the previous routine ended at time $t_0$, $r(t)$ is the entanglement distribution rate at time $t$, and $\Bar{r}(t)$ is the average rate of the period if we were to perform a compensation routine at time $t$.
Then the objective is to maximize
\begin{equation}
    \Bar{r}(t) = \frac{1}{(t-t_0)+c} \int_{t_0}^t r(t) dt
\end{equation}
A maximum must satisfy
\begin{align*}
    0&=\frac{d}{dt}\Bar{r}(t) \\
    &=\frac{(c + t - t_0) r(t) - \int_{t_0}^t r(t) dt}{(c + t - t_0)^2}
\end{align*}
which has a unique solution at
\begin{equation}\label{eqn:opt-derivation}
    r(t)=\frac{1}{c + t - t_0} \int_{t_0}^t r(t) dt = \Bar{r}(t)
\end{equation}

$\Bar{r}(t)$ is increasing at $t=0$, so assuming decreasing $r(t)$, which is generally true as we generally lower pump power and therefore $r(t)$ throughout a period, $\Bar{r}(t)$ is maximized by $t$ satisfying Equation~\ref{eqn:opt-derivation}.

\end{document}